# Short-Run Health Consequences of Retirement and Pension Benefits: Evidence from China☆

PLAMEN NIKOLOV[a,b,c,d†]        ALAN ADELMAN[a]

**Abstract:** This paper examines the impact of the New Rural Pension Scheme (NRPS) in China. Exploiting the staggered implementation of an NRPS policy expansion that began in 2009, we used a difference-in-difference approach to study the effects of the introduction of pension benefits on the health status, health behaviors, and healthcare utilization of rural Chinese adults age 60 and above. The results point to three main conclusions. First, in addition to improvements in self-reported health, older adults with access to the pension program experienced significant improvements in several important measures of health, including mobility, self-care, usual activities, and vision. Second, regarding the functional domains of mobility and self-care, we found that the females in the study group led in improvements over their male counterparts. Third, in our search for the mechanisms that drive positive retirement program results, we find evidence that changes in individual health behaviors, such as a reduction in drinking and smoking, and improved sleep habits, play an important role. Our findings point to the potential benefits of retirement programs resulting from social spillover effects. In addition, these programs may lessen the morbidity burden among the retired population. (*JEL* H55, H75, I10, I12, I19, J26)

**Keywords**: life-cycle, retirement, pension, health, aging, developing countries, China.

---

☆We thank the editor and the anonymous reviewers for comments that significantly improved the paper. We also thank Matthew Bonci, Dayne Feehan, Jake Tuckman, and Xu Wang for outstanding research support with this project. We thank David Canning, Subal Kumbhakar, Eric Edmonds, Susan Wolcott, Nusrat Jimi, Wei Xiao, Solomon Polachek, James MacKinnon and Morten Nielsen for constructive feedback and helpful comments. Plamen Nikolov gratefully acknowledges research support by The Harvard Institute for Quantitative Social Science, the Economics Department at the State University of New York (Binghamton) and the Research Foundation for SUNY at Binghamton. All remaining errors are our own. The authors declare that they have no conflict of interest.

†Corresponding Author: Plamen Nikolov, Department of Economics, State University of New York (Binghamton), Department of Economics, 4400 Vestal Parkway East, Binghamton, NY 13902, USA. Email: pnikolov@binghamton.edu

[a] State University of New York (Binghamton)
[b] Harvard Institute for Quantitative Social Science
[c] IZA Institute of Labor Economics
[d] Global Labor Organization

# 1. Introduction

Recent gains in longevity have contributed to the rapid growth of aging populations in many parts of the developing world. In developing countries, approximately 10% of the population is over the age of 60 as of 2018, and this number is projected to reach 30% by 2050 (Bloom, Canning, and Fink 2010). Developing countries face an additional economic challenge insofar as they must respond to societal changes in which a substantial portion of the population has become "old" before they have become "rich." In response to a combination of demographic, political, and economic concerns, governments in many developing countries have readjusted to the changing landscape by initiating new retirement policies and establishing innovative financing schemes for pension programs. The introduction of these programs will influence how and when individuals decide to retire. Moreover, it is likely that retirement programs will influence health expenditures because prolonged longevity is associated with increased morbidity rates during retirement years. Nonetheless, we know remarkably little about the consequences of these new pension programs.

In this paper, we examine the impact of China's New Rural Pension Scheme (NRPS) on various health outcomes, health behaviors, and health care utilization of Chinese adults in rural areas, particularly among adults age 60 and above. The NRPS program was introduced in 2009 in response to the rising age demographic and concerns about poverty among seniors (Holzman, Robalino, and Takayama 2009: 111–18).[1] In 2007, approximately 11% of China's population was age 60 or over, constituting 21% of the world's elderly population (UN 2007). Similar to other developing countries, the Chinese government faces additional pressing challenges: large rural and informal agricultural populations, high internal migration flows (Sabates-Wheeler and Koettl 2010), and weak institutions (Musalem and Ortiz 2011). The new program, a defined contribution pension program, was made available to all of the rural residents over 16 years of age.

Previous empirical research provided estimates regarding the influence of access to pension programs and retirement options on health outcomes; however, this research primarily relied on data from high-income countries. The majority of these studies point to retirement having a positive influence on various health outcomes.[2] Many studies report a significant increase in health after retirement (Charles 2004; Johnston and Lee 2009; Neuman 2008; Bound and Weidmann 2007; Coe and Lindeboom 2008; Coe and Zamarro 2011; Blake and Garrouste 2017; Latif 2013; Insler 2014). Other studies, though far fewer, detect significant negative effects on subjective health measures (e.g., Dave, Rashad, and Spasojevic 2008; Lindeboom, Portrait, and Van den Berg 2002; Behncke 2012; Sahlgren 2012).

---

[1] Feldstein and Liebman (2002) and Cutler and Johnson (2004) provide an historical overview of social pension programs in developed countries. Social pension programs are common in the developed world and are designed primarily to provide old-age insurance and aid consumption smoothing,

In this study, our primary focus is the impact of the NRPS program on adults age 60 and above because this group is eligible for immediate program benefits. The expansion of the policy affected an easily identifiable group because the policy was introduced in select geographic areas with variations across time and space. We exploit the staggered policy implementation by employing a difference-in-difference-in-differences (DDD) strategy. Using survey data from the Chinese Health and Retirement Study (CHARLS), we estimate intention-to-treat (ITT) effects by comparing the changes in health outcomes among individuals in those adults age 60 and above who were enrolled in the pension program to the changes in health outcomes among individuals in those adults age 60 and above who were not enrolled. We further augment the ITT analysis by instrumenting for the potentially endogenous enrollment decision. We instrument for individual NRPS enrollment status using the community NRPS enrollment status as we exploit identifying variation from the recent NRPS expansions at the community level.

We report three major findings. First, there is striking evidence that the NRPS program impacts direct measures of health. We detect strong positive effects on mobility: individuals living in areas covered by the plan reported a decrease in difficulty jogging 1 km, walking 100 m, and climbing a flight of stairs, reflecting an overall 4–13% decline in difficulty performing these three activities. We also detect a strong positive program impact on two additional proxies for the direct measure of health: vision and self-reported general health (Chatterji et al. 2002). We find that the program had no impact on affect, including emotional well-being and mental health outcomes.

Second, our empirical results show strong evidence that the NRPS had a positive impact on indirect measures of health. In fact, the program's beneficial impact on the domains of self-care

---

and customary daily activities was even greater than the program's effect on mobility. Individuals living in covered areas exhibited a 27–69% decrease in difficulty regarding self-care (getting dressed, eating, and bathing). They also experience decreased difficulty performing usual activities that are indirect measures of health, such as preparing meals, grocery shopping, and cleaning house.

Third, when we examine for potential mechanisms that could mediate the impact of the NRPS, we find that program access led to improvements related to behavioral inputs and better health care utilization.[3] Program participants reported benefits with respect to whether they drank regularly during the prior year (extensive margin) and their overall drinking frequency during the prior year (intensive margin). A large decrease in alcohol consumption was reported (significant at the 1% level). Program participants, on average, exhibited a 13% decrease in the reported frequency of alcohol consumption. In addition, our results show evidence that program participation reduced the incidence of cigarette smoking. NRPS participation had a positive effect on sleep patterns as well, and we detected a small positive impact on the number of hours slept (for the overall sample), although this finding was estimated imprecisely.

This paper makes three important contributions to the literature on pension programs in developing countries. First, this study is among the few studies to rigorously examine the effects of pension participation on individual health outcomes in the context of a developing country. Although recent studies have documented how pension participation and retirement affect health outcomes in the context of high-income countries (Insler 2014, Eibich 2015), only limited evidence exists regarding the influence of pension programs on health outcomes in developing countries. Our study focuses on China, the largest country in the world, with an overall population of 1.4 billion people. This focus on the most populated country in the world, by itself, showcases the importance of the issues from a public welfare standpoint. The setting of this social policy program is also unique because the NRPS program focuses on China's rural areas, where the economic dynamics resemble the economies of low-income countries. Second, we shed light on how program participation affects numerous health domains, and we examine impacts on health domains that are consistent with the World Health Organization's measurement of various direct and indirect proxies of health status (Chatterji et al. 2002). The work of Cheng et al. (2018) is most closely related to our study. Their study also explored the NRPS impact on some health measures, but it examined a narrower set of outcomes (self-reported general health and instrumental activities of daily living [IADLs]).[4] In contrast, we explore the program's impact on additional important and previously unexamined health domains using data from a more recent survey source. Specifically, we provide additional evidence of how NPRS participation influences mobility, affect, self-care, vision, social interaction, and various disease states.

---

[3] Only one other study in the context of a developing country (e.g. Cheng et al. 2018) attempts to examine how retirement influences health status through various channels. However, because of data limitations, that study was only able to examine channels related to nutrition, perceived healthcare access, and informal care. In contrast, and because of the rich data on various health behaviors, we are able to examine how the NPRS benefits influenced other important health behaviors, such as drinking, smoking, sleeping duration and actual health care utilization (as opposed to access only).

As health is a multi-dimensional concept, our study provides evidence of the benefits of program participation for additional health domains and we indirectly provide additional support for this important policy (Chatterji et al. 2002). Since the CHARLS survey includes numerous measurements within a specific health domain, we are able combine these specific measures into an aggregate index for each health domain, thereby providing robust evidence of program impacts that is less likely to be generated by multiple hypothesis testing.[5] Furthermore, a novel feature of our study is that the CHARLS survey we use is harmonized and adapted from the U.S. Health and Retirement Study (HRS) survey and other sister health surveys conducted in other high- and middle-income countries.[6] Therefore, harmonization of measures of health domains across surveys can enable future international comparisons of our findings (based on the CHARLS survey) with results from other studies that rely on other HRS sister retirement surveys.[7]

Our third contribution is that we provide evidence regarding important new channels that mediate the relationship between pension program access and improvements in health status. Specifically, we examine how program participation influences a wide array of health behaviors and time use. Due to the rich data on various health inputs, health access, and health care utilization, we can examine how program participation influenced health access and ascertain whether program access translated to better health care use. Our findings provide suggestive evidence that reduced drinking, reduced smoking (among females), and improvements in sleep duration likely play a very important role in driving positive results.[8]

---

[4] Cheng et al. (2018) examines cognitive function, psychological well-being, and mortality, whereas our primary focus is on mobility, affect, self-care, vision and social interaction, and various disease states. Nikolov and Adelman (2019) examine how pension program benefits impact intra-household inter vivos transfers.

[5] This approach follows Kling, Liebman and Katz (2007). Though the problems of multiple hypothesis testing have received some attention in economics, generally it has been overlooked.

[6] The Health and Retirement Study Survey Research Center provides extensive information on a growing network of longitudinal aging sister studies (e.g., the Health and Retirement Study in the U.S., the Survey of Health, Ageing and Retirement in Europe, the English Longitudinal Study of Ageing, etc.) around the world.

[7] The U.S. Health and Retirement Survey (HRS) has become the model for a growing network of longitudinal aging studies around the world. These surveys not only provide data for individual countries but also offer the opportunity for cross-national comparisons. More information on the harmonization of variables across various global aging surveys is available at: https://hrs.isr.umich.edu/about/international-sister-studies

The remainder of this paper is structured as follows. Section 2 provides background information regarding rural pension programs in China and the NRPS program. Section 3 summarizes the data. Section 4 presents our identification strategy. In Section 5, we provide a discussion of the results. Section 6 presents various robustness checks, and Section 7 concludes the paper.

## 2. China's New Rural Pension Scheme
### 2.1 Background

China introduced rural pension schemes in 1986, beginning with a pilot program for rural residents before expanding coverage further. The program financing relied on voluntary contributions from individuals that were matched by contributions from local governments. By the end of 1998, two-thirds of the rural counties were covered, amounting to 2,123 counties in 31 provinces. However, a combination of poor governance, unsound local operations, and inflationary pressures brought about by the Asian financial crisis in 1997, halted the rural pension expansion. In 1999, the program was scaled down following concerns about its long-run sustainability in rural areas. Pension coverage fell from 80.25 million participants in 1998 to roughly 55 million among rural participants in 2007.[9]

The New Rural Pension Scheme (NRPS) was launched in 2009. The program aimed to achieve full geographic coverage by 2020 (Dorfman et al. 2013; Cai et al. 2012). The program offers a basic flat pension financed by the central government, individual contributions, and a minimum matching contribution from local governments. The program covered 23% of China's counties by the end of 2010 and over 60% of counties by early 2012. Figure 1 shows the expansion of the program's coverage over time.

---

[8] Studies by Insler (2014) and Eibich (2015), conducted with data from developed countries, investigated health effects and potential mechanisms.
[9] The rural pension system faced continuing challenges in the early 2000s. Program participation favored wealthier regions and poor provinces failed to make their matching contributions. The program witnessed a resurgence from 2003 as interest grew and individual participation rapidly increased. More than 300 counties and 25 provinces introduced program benefits by the end of 2008 (Dorfman et al. 2013).

[Figure 1 about here]

Total participation grew to 326 million from 2009 to the end of 2011 (Quan 2012) and over 50% of rural residents participated in the NRPS by the end of 2011. Three important factors likely account for this program expansion. First, China's dedication to rural pension reform and the country's high economic growth rate from 2009–2011 played a role. Second, the uneven geographic coverage generated demand for the basic monthly benefit in untargeted areas. Third, the rural pension expansion was a key element in the domestic political and policy discourse in 2012, and the political debate further intensified interest in the program.

## 2.2    NRPS Program Eligibility and Benefits

The NRPS is available to all of the rural residents over age 16 who are not enrolled in an urban pension program. The program was introduced in the rural administrative districts called *Hukou*. Program participation is voluntary, and those who contribute for at least 15 years are eligible to receive benefits at the age of 60. Rural residents who are over the age of 60 at the start of the program are eligible to receive the basic monthly benefit of 55 RMB if their children already contribute to the pension scheme.[10] Participants between the ages of 45–60 with less than 15 years of contributions are encouraged to increase monthly payments to cover the absence of contributions during their earlier working years.

Individual contributions are voluntary and range annually from 100 to 500 RMB (approximately 15–77 US dollars). Based on a 2009 survey, the mean participant contribution was 100 RMB (Dorfman et al. 2013). The local governments are required to match 30 RMB annually per individual contribution. Participants between the ages of 45–60 with fewer than 15 years of contributions are encouraged to increase monthly payments to cover the lack of contributions during their work cycles. Rural residents who opt to participate in the program must register at their local village government office, the lowest level of the Chinese government hierarchy. Payments are made at that office in cash throughout the year. At the end of each monthly contribution period, each village government aggregates the payments received and transfers that money to the next level of government (a township), which then transfers the contributions to the next level (the administrative county). Although there has been some discussion that the funds should be managed at the province level, currently the rural pension funds generally are managed by the county-level government. In general, the contributions are deposited in banks, and administrative expenses are picked up by the local governments.

---

[10] The central government fully subsidizes the basic pension in Central and Western provinces and splits the cost with local governments in Eastern provinces (Cai et al.2012).

## 3.    Data Description

### 3.1    Survey Data

*China Health and Retirement Longitudinal Study*. Our primary data source for this study was the China Health and Retirement Longitudinal Study (CHARLS) from which we drew data about retirement status, pension program access, health status, health care use, health behaviors, and socio-economic information for the individuals surveyed. CHARLS is a nationally representative survey that sampled individuals 45 years of age or older. The total sample comprised 17,708 individuals living in 10,287 households in 450 villages/urban communities in 150 counties/districts across 28 of China's 30 provinces, excluding Tibet. The 2011 baseline wave of data for CHARLS was gathered from interviews of 10,257 households with 18,245 respondents age 45 and over.[11] The 2013 CHARLS survey covered 10,979 households with 19,666 respondents. In the 2013 wave, the interviewers followed up with 88.6% of the original respondents and 89.6% of original households,[12] and added 2,053 new households with 3,507 individuals. Figure 2 shows the survey coverage map.

CHARLS collected demographic information along with data regarding family structure, subjective and objective health status, health care use, pensions and retirement status, work, household wealth, income, and consumption. Basic information was collected for each individual, couple, and household regarding their gender, age, education, household size, and marital status. The survey response rate was over 80% (94% in rural areas and 69% in urban areas).[13]

---

[11] Initially, 19,081 households were sampled where 12,740 had age-eligible members, of which, 10,257 responded.
[12] 16,159 of the original 18,245 respondents and 9,185 of the original 10,257 households.

We use data from the CHARLS regarding individual NRPS participation, individual socio-economic characteristics, health outcomes, and other community-level data. Table 1 presents summary statistics for several characteristics of the respondents. Among the sample of individuals eligible for NRPS participants, 70% were employed in 2011, while 69% of non-participants were employed that same year. About three-fourths of the total sample worked in agriculture: 72% of the participants and 73% of the non-participants. We use the geographic variables in the data to identify the rural subsample. Our rural sample reported low levels of educational attainment: only approximately 46–48% reported having completed at least a secondary level of education. In terms of self-reported health status, approximately one-fourth of both the participants and non-participants reported being in "at least good health." For health behaviors, we find that one-third of the participants and non-participants smoked cigarettes and drank alcohol at the time of the survey. In the baseline survey, both participants and non-participants reported sleeping an average of 6.4 hours daily. The means for the categories under health care utilization were also similar between the participants and non-participants in the base year. We see that 10% of participants and 9% of non-participants reported staying in the hospital overnight during the prior year. With respect to the incidence of disease states, hypertension was more prevalent than diabetes in our sample: 25% versus 5% of the sample among participants, respectively, and 22% versus 4% of the total sample (participants and non-participants alike), respectively.

[Table 1 about here]

*China Health and Nutrition Survey.* Our secondary data source is the China Health and Nutrition Survey (CHNS) from which we draw data prior to 2009 regarding health status, health care use, health behaviors, and individual and county level socio-economic information. The CHNS survey enables us to examine data prior to the introduction of the NRPS. The CHNS is a longitudinal survey that covered about 19,000 individuals in 15 provinces spanning 216 primary sampling units (PSUs).[14] Figure 2 provides the coverage map for the CHNS survey.

---

[13] The survey sampling occurred in three stages. First, all of the community-level units, were stratified into 8 regions, by rural community and urban districts, and by community/district GDP per capita. After this step, 150 counties were randomly chosen using probabilities proportional to size (PPS). Within the 150 counties, 3 primary sampling units (PSUs) were selected randomly using the same PPS method.

[Figure 2 about here]

The first wave of the survey started in 1989 and aimed to provide data on how the economy and social factors impact individual health and nutritional status. The panel survey covered years 1989, 1991, 1993, 1997, 2000, 2004, 2006, 2009, and 2011. The survey modules covered food choices, nutritional intake, health behaviors, physical activities, work activities, time usage, and nutritional status. The sample was selected using a multi-stage random selection process, which was similar across panel waves. First, counties were stratified by income level, followed by a weighting scheme that selected 4 counties from each province (CHNS Research Team 2010).

The main component of the CHNS survey is the individual module. From the 2004 survey onwards, all of the questions relate to individual activities, lifestyle, health status, demographic status, body shape, mass media exposure, and other activities were part of the individual questionnaires. There are two parts of the individual questionnaire: one for adults aged 18 and older, and the second part for children under age 18. Children age 6 and above and all of the adults provided information about how they allocated their time between household and physical activities, and about their food and beverage consumption. For individuals 12 years and older and for all of the adults, additional questions covered smoking status, alcohol consumption, diet, and physical activity. Adolescents age 12 and older and women under age 52 with children age 6–18 in the household were asked additional questions related to mass-media exposure. Adults age 55 and older were asked to provide information about their daily living activities and were given a memory test.

## 3.2    Health Outcomes

For this paper, we follow the typology of health domains provided by Chatterji et al. (2002) and we employ both direct and indirect measures of health, supported by the data from CHARLS.

---

[14] The survey covered the following provinces (also presented in Figure 2): Bejing, Chongqing, Guangxi, Heilongjiang, Henan, Hubei, Hunan, Jiangsu, Liaoning, Shaanxi, Shandong, Shanghai, Yunnan, and Zhejiang.

For direct measures of health, we focus our attention on self-reported health[15], mobility, affect, and vision. Each survey respondent was asked a series of questions about his/her functional limitations and health status. For mobility, a direct health measure that refers to an individual's difficulty with movement, the questions related to difficulty jogging for 1 km, walking for 100 m, and climbing a flight of stairs. The questions were coded dichotomously, where reporting some difficulty or an inability to complete the task was set to 1. The affect domain included questions about the respondent's reactions in situations that measured mood, outlook on life, and levels of distress, sadness, and anger. The CHARLS survey asked respondents 10 mental health questions pertaining to the week prior to the interview. We chose 3 main statements: "I felt everything I did was an effort," "I was happy," and "I felt lonely." The respondents were asked to select from four choices: "rarely (< 1 day)," "some (1–2 days)," "occasionally (at least 3 days)," and "mostly (5–7 days)." Two questions were coded to 1 if "mostly" and 0 if "otherwise," whereas the happiness variable was equal to 1 if rarely happy. Vision was the last direct measure of health in our focus. Respondents were asked to rate their vision on a 5-point scale[16] based on their self-assessed ability to see short and long distances. The questions were coded dichotomously to equal 1 if the respondent rated his/her vision "at least good."

To be consistent with the WHO definition of health domains (Chatterji et al. 2002), we also examine for program impacts on indirect measures of health. To measure health indirectly, we use data on self-care, usual activities, and social interactions. The self-care domain reflects the ability to take care of oneself daily.[17] We rely on questions about daily living, including difficulty getting dressed, taking a bath, and eating a meal (cutting up food), all of which are used for the self-care index. To measure the impact of retirement on the usual activities domain, we rely on a set of questions regarding daily activities. CHARLS asked respondents about difficulty preparing hot meals, shopping for groceries, and cleaning the house. These questions assessed a person's ability to perform routine tasks properly. If the respondent reported some difficulty or an inability to complete the task, these indicators were coded as 1. A final indirect health domain within our focus is social functioning. Respondents were asked about their contact with parents, in-laws, and children. We use data for three questions regarding any weekly contact with parents or in-laws, any weekly contact with children by phone or email, and any weekly contact with children in person or by phone or email. These variables were set to 1 if there was contact in the past year at least once per week.

---

[15] Self-reported health is strongly predictive of life expectancy among adults (Idler and Kasl 1995; Idler and Benyamini 1997). We measured self-reported health by asking "In general, how is your health: excellent, very good, good, fair, or poor?" Respondents were asked to rate their health on a 5-point scale. We coded 1 if self-reported health was at least good, and 0 otherwise.
[16] The scale range was (1) excellent (2) very good (3) good (4) fair, and (5) poor.
[17] Many national surveys use measures of activities of daily living (ADLs) to assess an individual's functional status, self-care, and quality of life (Wiener et al. 1990).

Since we have multiple outcomes for each health domain, we use principal component analysis to create a standardized aggregate index for each domain.[18, 19] The indices for mobility, affect, self-care, and usual activities were coded so that high values were associated with difficulties or negative feelings.[20] Low (or negative) values denoted less difficulty and better mood or emotions. Vision and social interaction indices were coded positively, and high values were associated with positive social interaction and better vision status.

In addition, we examine for mechanisms that might mediate impacts on various health domains by using survey data on disease states, health care utilization, and health behaviors. Examining changes in such mediating factors can enable better understanding of how these changes affect health status directly or indirectly. Various individual behaviors, such as smoking, alcohol consumption, BMI, and sleep duration, are inputs to the production function of health. Healthcare utilization (e.g., doctor visits) is one way to improve the state of health through direct treatment, and the lack of inpatient or outpatient treatment can have adverse effects on one's health. We also analyze morbidity changes by using the incidence of hypertension and diabetes. Although these measures do not capture the level of health, they are inputs to one's health status that subsequently can influence the measure of a particular health domain.

CHARLS collected detailed information on each participant's smoking, alcohol consumption, and sleep duration. We use an indicator of current smoking status and the self-reported number of cigarettes smoked daily. For alcohol consumption, we rely on three survey questions with corresponding indicators for consuming alcohol during the past year, consuming alcohol at least once a week in the past year, and the frequency of consumption in the past year. The frequency is coded by categories: (0) none or doesn't drink; (1) once a month; (2) 2 to 3 days a month; (3) once a week; (4) 2 to 3 days a week; (5) 4 to 6 days a week; (6) daily; (7) twice a day; (8) more than twice a day. We use data on each participant's weight and average daily number of hours of sleep per night over the past year. Using each respondent's height and weight, we calculate the respondent's BMI. We use the respondent's BMI to code an indicator equal to 1 if the reported BMI is less than 18.5. We also use data on health care utilization for hospital visits and outpatient doctor visits.[21]

---

[18] With the use of Principal Component Analysis (PCA), we reduce each multi-dimensional domain into a composite index. Principal Component Analysis (PCA) is used to reduce the dimensionality of correlated variables that measure essentially the same domain. This is achieved by estimating $n$ weighted linear combinations (reported in Online Appendix Table A.1) containing the proxy variables within each domain. The linear transformation produces $n$ uncorrelated components (linear combinations) that are the eigenvectors of the system; combined they contain the same information as the original variables. By design, the first component contains the most information (largest eigenvalue), whereas the last component contains the least. We reduce the multiple dimensions into one, retaining the component with the largest overall variance (eigenvalue).

[19] We use the correlation matrix in the weighting procedure, where the weights (loadings) are equal and not sensitive to the scale of the variable. Although the data is commensurable, we avoid assuming multivariate normal. Also, the package computes standard errors as if it were the covariance matrix, even when using the correlation matrix.

[20] The weight for difficulty getting dressed enters positively in the linear equation for the composite score of the self-care index. This suggests that more difficulty within this first component translates into high levels of the self-care index, i.e., increased difficulty in self-care tasks overall. The self-care index has a mean of 0 and a standard deviation of 1.42. We repeat the procedure for the remaining domains that have multiple dimensions. Online Appendix A provides the loading tables for all of the indices.

## 4.    Identification Strategy

Our basic empirical strategy is the DDD approach as we estimate the program's impact on various health domains by comparing changes in health outcomes for adults age 60 and above who live in treated areas to changes in health outcomes for adults age 60 and above who live in areas that were not offering the NRPS program during the study period.[22]

Therefore, our identification strategy relies on the timing differences among communities for pension policy adoption. By the end of 2010, 23% of all communities were covered and over 60% were covered by 2012 (Dorfman et al., 2013; Cai et al., 2012), as shown in Figure 1.[23] We use this staggered implementation as a source of identifying variation to distinguish between the effects of pension participation on the health outcomes of individuals age 60 and above living in communities that started offering benefits between 2011–2013 and the health outcomes of adults age 60 and above living in areas that were not offering the pension program. We construct $OfferNRPS_{ct}$ for communities that offered the NRPS program in time $t$. Given data limitations, we construct OfferNRPS$_{ct}$ based on individual-level data. If no individuals indicate having NRPS at time $t$ in community $c$, then OfferNRPS$_{ct}$ equals 0. If at least one person reports participating in the NRPS, then *OfferNRPS$_{ct}$* is set to 1.[24] We interact *OfferNRPS$_{ct}$* with a binary indicator for age 60 or over, following the empirical approach in Katz (1996), Gruber (1994), and Rossin (2011).

---

[21] Hospital visits were collected for the year prior, whereas doctor visits were surveyed for the month prior to the interview. The inpatient indicator equals 1 if the respondent stayed overnight at a hospital at least once in the past year. The outpatient variable is equal to 1 if the respondent visited a doctor at least once in the past month. Other measures of utilization are out-of-pocket (OOP) expenditures. Inpatient OOP expenditures exclude the amount paid by insurance, the wages paid to a hired nurse, transportation costs, and accommodation costs for the respondent and family members. Outpatient OOP expenditures exclude the amount paid by insurance, but include the fees paid for treatment, medication costs, and prescription drugs.
[22] As opposed to difference-in-differences (DD), the DDD estimate is robust to several potential confounders: changes among the elderly in control communities and changes among individuals under the age of 60 in treated communities (assumed to be unaffected in time t).
[23] Demographic information in the CHARLS survey is only available at the community (*shequ*) level.

We estimate:

$$(1) \qquad Y_{ict} = \beta_0 + \beta_1(OfferNRPS_{ct} \times Above60_{ict}) + \beta_2 Above60_{ict}$$
$$+ \beta_3 \boldsymbol{X}_{ict} + \phi_c + \mu_t + \phi_c \times \mu_t + \varepsilon_{ict},$$

where Y$_{ict}$ is the health outcome studied and *Above60$_{ict}$* is equal to 1 if the respondent is age 60 and over. $\beta_1$ in (1) is the coefficient that captures the estimate of the average effect of program availability on the average outcomes of eligible individuals age 60 and over who live in a treated community, regardless of whether they decided to participate in the program (i.e., the ITT effect). $\boldsymbol{X}_{ict}$, is a vector of individual-level controls, education, gender, age, age squared, household size, and marital status. $\phi_c$ and $\mu_t$ are community-level and time fixed effects, respectively. Community-level fixed effects allow us to control for time-invariant characteristics that affect the likelihood of program availability in the community. Year fixed effects control for characteristics common across time in communities.[25,26] In addition, we use community-time fixed effects, $\phi_c \times \mu_t$, to control for community differences during the implementation of the NRPS.

The identification assumption is that treatment communities that provide program benefits would otherwise have changed in a manner similar, on average, to the control communities that did not provide program benefits. Even though the identification assumption cannot be tested directly, in practice, by controlling for district-by-year fixed effects, to some extent, we can examine whether the two groups (treated and non-treated) exhibit parallel trends in the outcomes prior to 2009 when the program was introduced. Since all of the survey data from CHARLS is post-NRPS program, we use data from the CHNS survey to compare pre-trends of study outcomes.

---

[24] We address potential concerns regarding measurement error and associated bias in the estimated coefficients based on this approach in Online Appendix B.
[25] We cluster the standard errors by community and age groups based on Bertrand, Duflo, and Mullainathan (2003).
[26] In Online Appendix B, we report additional robustness checks where we cluster the standard errors by community and age. We show that our results are robust to community and age specific clusters.

We conduct a formal test to examine the possibility of differential trajectories between treated and non-treated areas based on the work of Autor (2003).[27] We use CHNS survey data prior to 2009 (when the NRPS program coverage began) for the various health domains and health behaviors from the survey's 2000, 2004, and 2006 waves. Specifically, we estimate:

$$(2) \qquad Y_{ict} = \beta_0 + \beta_{-3}D_{ct} + \beta_{-1}D_{ct} + \phi_c + \mu_t + \varepsilon_{ict} \,,$$

where $Y_{ict}$ is the health outcome studied, and $\phi_c$ and $\mu_t$ are community-level and time fixed effects, respectively. We include the interactions $D_{ct}$ of the time-period dummy variables and the treatment indicators for the first pre-treatment period and last pre-treatment period.[28] The results reported in online Appendix Table A.2 provide clear evidence that $\beta_{-3}$ and $\beta_{-1}$ are insignificant. Therefore, we cannot reject the hypothesis that trends in the outcomes between the treatment and control areas are the same. In other words, based on the data from the CHNS survey, we see no evidence of statistically significant differences between the two groups in the pre-treatment period.

In an extension to specification (1) above, we also examine the impact among individuals who actually participated in the program. We use $OfferNRPS_{ct}$ to instrument individual participation in the NRPS Program. $OfferNRPS_{ct}$ is set to 0 if no individuals participate in the NRPS, and it is set to 1 if the community has at least 1 participant. We estimate:

$$(3) \qquad Y_{ict} = \beta_0 + \beta_1(\widehat{NRPS}_{ict} \times Above60_{ict}) + \beta_2 Above60_{ict}$$
$$+ \beta_3 \boldsymbol{X}_{ict} + \phi_c + \mu_t + \phi_c \times \mu_t + \varepsilon_{ict}.$$

---

[27] Online Appendix Figure A.1 shows trends for the direct and indirect measures of health and health behavior outcomes for the years prior to the introduction of the NRPS program. Each figure displays outcome averages from the 2000, 2004, and 2006 waves of the CHNS for treated and non-treated areas. The control provinces are Hunan and Guizhou, which were identified based on our CHARLS sample of communities as not offering the program from 2011 to 2013. The figures provide no visible support for differential trends between treated and non-treated areas.

[28] We omit the interaction for the second pre-treatment period.

$\widehat{NRPS}_{ict}$ represents individual enrollment in NRPS, and we instrument it with $OfferNRPS_{ct}$. $X_{ict}$ is a vector of individual-level controls, and $\phi_c$, $\mu_t$, and $\phi_c \times \mu_t$ are community-level, time, and community-time fixed effects, respectively.

# 5.    Results

## 5.1    Pension Program Impacts

First, we present the intention-to-treat effects based on specification (1). Table 2 (Panels A and B) reports results for the direct measures of health: self-reported health, mobility, affect, and vision. Table 3 reports the results for the indirect measures of health: self-care, usual activities, and social interaction. Table 4 reports the results for the disease states.

[Table 2 about here]

[Table 3 about here]

The results provide striking evidence of the positive impact of the NRPS program on health domains involved in daily functioning. We detect strong positive effects on mobility (reported in Table 2) and on self-care and usual activities (reported in Table 3) among individuals living in NRPS program areas. For example, individuals living in treatment areas reported decreased difficulty jogging 1 km, walking 100 m, and climbing a flight of stairs, for an overall decline in difficulty on these three outcomes of 4–13%. The program also had a negative impact on the overall PCA-based mobility index of approximately 0.11 units (0.08 standard deviations). We find a positive program impact on self-reported health, although the effect was not statistically significant at the 0.10 level. The pattern of positive impacts persists for the indirect measures reported in Table 3. In fact, the program had a greater impact on the indirect measures used as proxies for the self-care domain. Individuals living in treated areas exhibited a 27–69% decrease in difficulty with self-care measures (getting dressed, eating, and taking a bath). This improvement was reflected in the program's impact on the overall PCA-based self-care index, which also exhibited a sizable decline. The program improved self-care functioning by 0.15 units, or 0.10 standard deviations. We detect beneficial program impacts on affect and vision, where the program improved functioning in the vision domain by 0.09 (0.08 standard deviations). This result is consistent with the findings based on pension programs in high-income countries. For example, Coe and Zamarro (2011), Johnston and Lee (2009), and Eibich (2015) report beneficial impacts on indirect proxies of health that were statistically significant.

Table 3 also reports the program's beneficial impact on the domain of usual activities. Individuals living in program areas experienced a 14 –20% decrease in difficulty in the performance of the proxies for usual activities (preparing meals, grocery shopping, and cleaning house). We detect a statistically significant decrease in the overall PCA index of usual activities of 0.11 units (0.07 standard deviations). We detect a program impact, although not statistically significant, on the social interaction index, which captures contacts with parents and with children.

In Table 4, we report the impact on the incidence of disease states and morbidity among participants. We examine disease states separately because health status outcomes must be understood separately from diagnoses. Disease states with the same diagnosis may be associated with very different levels of health in different individuals, or in the same individual at different points during the course of the disease. For example, a person with diabetes in the initial course of the illness may be fully functional with just some dietary restrictions and exercise prescribed. However, diabetes may progress to renal failure or retinopathy that severely limits the person's functioning. In other words, information beyond diagnosis is critical for understanding health at the individual level. In Table 4, we do not detect a program impact on the incidence of recent diagnoses of diabetes or new diagnoses of hypertension.

[Table 4 about here]

Tables 2–4 also present estimates for the difference-in-difference strategy combined with the instrumental variable approach based on (2). Table 2, Panel B, displays the two-stage least squares results (LATE) on direct measures of health; Table 3 Panel B reports the results for the indirect measures of health; and Table 4 Panel B reports the instrumental variable results for program impacts on two disease states. For all of the outcomes, the tables report very high F-statistics (well above 10).

The local average treatment effect (LATE) results reported in Table 2, Panel B, provide evidence consistent with strong positive program ITT impacts on the for functioning-related health domains. Program participation led to strong positive effects with regard to mobility, self-care, and usual activities (Table 3 Panel B).[29] Program participants experienced a 9% decline in difficulty jogging 1 km, a 16% decline in difficulty walking 100 m, and a 29% decrease in difficult climbing a flight of stairs. The NPRS program also had a negative impact on the overall PCA-based index of 0.24 units, or 0.18 standard deviations. Among the participants who chose to participate in the program, we found a positive impact on self-reported health, although it is not statistically significant at the 0.10 level. We do not detect statistically significant effects on the affect domain. Only the effect on the overall index for the vision domain is marginally statistically significant at the 0.10 level.[30]

In terms of indirect measures (reported in Table 3 Panel B) for the *LATE* results, we find that program participation leads to a 59–152% decline in difficulty with self-care measures (getting dressed, eating, and taking a bath). This improvement is reflected in the program's impact on the overall PCA-based self-care index, which also exhibited a very sizable decline of 0.32 units (0.22 standard deviations). Finally, the results in Table 4 Panel B reveal no impact among program participants on the incidence rates of diabetes and hypertension. It is likely that this finding is a byproduct of the short time window of the study, which may have prevented us from seeing results regarding these incidence rates.

The previous analysis focuses on changes resulting from the offering of NRPS coverage and program participation because data on program benefits across different areas relies on administrative data. In contrast, the retirement variable indicating actual retirement relies on self-reported data, and the retirement status variable itself is sparsely distributed. Using data on self-reported retirement, we repeat the combined difference-in-difference strategy with the instrumental variable method based on (2), although we examine for program impacts by comparing the retirement decision with the offer of participating in the NRPS program. Table 5 reports the results.

---

[29] Positive effects on the functioning of health domains also means reduced difficulty shown by these PCA indices and individual survey responses.
[30] Coe and Zamarro (2011), Johnston and Lee (2009), Eibich (2015) found positive effects on physical health, although the results were significant only in Coe and Zamarro (2011). For mental health, Coe and Zamarro (2011) detected no statistically significant effects, while Johnston and Lee (2009) and Eibich (2015) reported beneficial impacts that were statistically significant.

[Table 5 about here]

These findings corroborate the patterns of the main *LATE* results: pension program participation confers benefits in three important functioning domains: mobility, self-care, and usual activities(imprecisely estimated for the domain of usual activities). The results on the composite indices for all the self-care domains is significant at the 0.10 level.

## 5.2    Mechanisms

NRPS beneficiaries likely experienced a whole host of behavioral changes when they obtained access to the NRPS benefits, and that could account for the impact of the program on various health domains. To better understand the channels underlying the influence of the NRPS, we examine how program participation affected various health behaviors, time use, and health care utilization. If we observe no change in a measure that could play a mediating role on the health outcomes studied, then we would have a strong indication that the causal pathway does not operate via that mediating factor. In particular, two factors could change among program participants: time allocation and behavioral adjustments. Due to data limitations, we are unable to explore time use (exception for sleep duration), but we focus on several health outcomes, as reported in Table 4.

We examine the program's impact on participants' reported incidence of alcohol use during the prior year, including regular drinking and the overall frequency of drinking. We detect a very large decline in alcohol consumption (statistically significant at the 0.01 level) among program beneficiaries. Program participants, on average, exhibited a 13% decrease in the reported frequency of alcohol consumption. In addition, we examine program effects on sleep patterns, and we detected a small but positive impact on hours slept. Although not quite significant (at the 10% level) for the overall sample, this finding is significant (at the 10% level) for the male sample and with a larger effect size. We see some evidence of a decrease in the reported prevalence of individuals being underweight, although it is not significant at the 10% level. We compare the pattern of these beneficial program impacts on health-related behaviors, and we note that the observed changes are similar to changes reported by retirement programs in high-income countries (Insler 2014; Eibich 2015).

Next, we turn our attention to health care access and health care utilization among program beneficiaries. We detect no impact on the ITT estimates for inpatient or outpatient visits or out-of-pocket health expenditures. We also do not detect program impact on the outcomes for the two disease states for which we had data: incidence of diagnoses of diabetes or hypertension in the past year. In sum, Table 4 provides suggestive evidence that various mediating factors—including improved health behaviors, such as reduced smoking, reduced alcohol consumption, and improved sleep duration—played a positive role in influencing the health domains. We do not detect an impact on health care access, health care utilization, or specific disease states for which we had data.

## 5.3    Heterogeneous Treatment Effects

We also examine whether the impact of program participation differed by gender. To identify any differences of program benefits based on the gender of the participants, we repeat specifications (1) and (2) for the male and female samples separately. Table 6 reports the ITT and LATE results.

[Table 6 about here]

Table 6 provides evidence that the onset of program eligibility did confer different health benefits by gender. In terms of the direct measures of health, the overall pattern is consistent across genders, except that women also reported improvements in the affect domain, and these improvements were statistically significant at the 10% level. In terms of another direct measure of health, females reported much larger improvements in mobility than reported by men, as exhibited by the statistically significant effect for the female sample. Turning our attention to the indirect measures of health, we observe that although both males and females over 60 exhibited stark improvements in self-care and usual activities, the effect sizes were much larger for females than for males. The results on health behaviors, health care access, and health care utilization also reveal important gender differences. Male program beneficiaries reported decreased alcohol consumption and decreased drinking frequency. Females beneficiaries reported a large decrease in the incidence of current smoking.

# 6.    Robustness Checks

Next, we present a number of consistency checks to investigate whether the magnitude of our estimates are within a reasonable range. We also provide additional robustness checks to address possible identification threats related to measurement issues or specification choices.

## 6.1    Falsification Tests

To test against potential spurious effects, we construct a falsification exercise based on specification (1). The NRPS program is available to individuals living in rural administrative districts so long as these individuals are not already enrolled in an urban pension scheme. In our main analysis, we use survey responses to remove urban pensioners who are not eligible for the NRPS program) but who happen to live in rural administrative districts (rural *Hukous*). However, for this falsification exercise, we reconstruct a sample that only includes these respondents in the falsification sample and we re-run specification (1) but only using urban pensioners. The goal of this exercise was to see whether there was a change in health outcomes between urban pensioners age 60 and above who live in NRPS-treated areas versus the health outcomes of urban pensioners ages 60 and above who live in non-treated areas. Since urban pensioners are not allowed to participate in the NRPS program and not directly affected by the NRPS program coverage, we should not see significant program impacts on the health outcomes for this group of individuals based on specification (1).

Online Appendix Table A.3 reports the results and provides evidence that $\beta_3$ is not significant for any of the outcomes. As predicted, we detect no evidence of program effects among the urban pensioners.[31]

---

[31] As an additional falsification exercise, we also re-estimated (1) and (2) on a set of placebo outcomes for which there is no conceptual basis to detect program impacts. We performed this additional robustness check to provide additional evidence that the results were not driven by the specification choice. Online Appendix Table A.4 presents the results of four outcomes for which there is absolutely no conceptual basis to expect program impacts: the likelihood of one's nationality being Han, the number of female household members, the number of daughters in the

## 6.2 Alternative Proxies of Health Status Outcomes

To further examine the robustness of our results, we use additional measures of health available from the CHARLS survey. The survey collected additional secondary proxies for the domains of self-care, usual activities, mobility, and affect, which are the main health domains for which we detected positive program impacts. We examine the program impact on these additional proxies to bolster the credibility of our main results. Specifically, we use survey questions about the respondent's capacity to get in and out of bed, use the toilet, manage money, take medications, kneel or stoop, and stand after sitting for a long time.[32] We also examine two variables that measure the affect domain: being bothered by little things and feeling depressed.[33] Appendix Table A.5 reports the ITT and LATE estimates using (1) and (2). As reported in Appendix Table A.5, the results confirm the pattern of beneficial program impacts on the domains of mobility, self-care, and usual activities.

## 6.3 Alternative Measures of NRPS Participation

Measurement error in self-reported information regarding NRPS program participation could bias our estimates. To bolster the credibility of our estimates, we address this potential concern with additional robustness checks based on alternative measurements of NRPS participation.

*Propensity Score Definition*. First, to deal with possible mismeasurement of NRPS participation based on the survey response, we re-construct the individual participation status using a propensity score matching approach. Specifically, we predict NRPS participation for each wave of the CHARLS survey using a linear combination of the respondent's education, gender, parental education, and nationality (set to 1 if the respondent was of Han nationality) with information for each variable at baseline in the CHARLS survey. We use the predicted propensity of NRPS participation, $\widehat{NRPS}_{ic}$, which is based on the propensity score matching method, to define a new

---

household, and the mother's educational level. Panel A reports the ITT results, and Panel B reports the LATE results. Both panels report no evidence of program impact on the four outcomes, which bolsters the validity of the results.

[32] If the respondent reported some difficulty or an inability to complete the task, we coded these indicators as 1.

[33] These variables equal 1 if the respondent reported having felt those emotions at least 5–7 days weekly prior to the interview.

variable, $PrNRPS_{ic}$. Based on the propensity score, we construct a binary variable, defining $PrNRPS_{ic} = 1$ if $\widehat{NRPS}_{ic}$ is greater than 1 standard deviation above the mean $\widehat{NRPS}_{ic}$. We then proceed with the DDD analysis as in the main portion of our analysis, except that program participation in this analysis is based on the newly reconstructed variable ($PrNRPS_{ic}$) for NRPS participation status. This participation variable was based on the propensity score estimation rather than the self-reported variable from the CHARLS survey. The results, as reported in Online Appendix Tables B.1, B.2, and B.3, demonstrate that our findings are robust to this alternative definition of program participation.

*Community NRPS Participation Definition.* Next, we consider the possibility that there can be a measurement error resulting from noisy self-reported data for the NRPS program participation variable. The measurement error for the individual NRPS participation variable can creep into the variable that we use to capture the community-level NRPS participation.[34] Therefore, we check the robustness of our definition of $OfferNRPS_{ct}$ to correct for possible contamination of the treated group resulting from random noise in reporting at the individual level. To address this potential concern, we raise the threshold when *OfferNRPS$_{ct}$* (the community participation variable) is set to 1. Instead of setting the threshold to at least 1 individual reporting NRPS participation to set $OfferNRPS_{ct}$=1, we use an alternative higher threshold of at least 4 (or at least 7 for an even higher threshold) participants in community *c* to set $OfferNRPS_{ct}$=1. We report the results from both additional analyses in Online Appendix Tables B.7, B.8, and B.9. The results show that our original estimates are robust to alternative definitions of the threshold defining $OfferNRPS_{ct}$.

*Using Online Administrative Data.* Finally, we bolster the credibility of the direction of our estimated program impacts by performing a final robustness check. For this exercise, we use administrative data available online based on public announcements regarding which cities—and the list of communities within these cities—that definitively participated in the NRPS program in Heilongjiang Province. Specifically, based on online newspaper announcements, we were able to identify the timing of when specific cities (and communities within these cities) switched from being non-participants to being participants in the NRPS program. However, we faced a challenge because we could not map the actual communities (within cities) to the communities listed in the CHARLS survey, which we used in the main analysis, because the community definition in the CHARLS survey differs from the actual administrative *community unit*. Although we could not re-do the previous analysis (done at the community level ) and replace the CHARLS survey data we previously relied on with admin data for community level participation, we could re-run our previous specifications at the city level because: (1) we knew the number of known communities within each city that definitively participated in the NRPS program based on the online public announcements, and (2) we knew the total number of all of the communities within each city and that number is fixed. We re-defined our treatment variable from the binary one used in our main analyses (at the community level) to treatment intensity (at the city level)[35] and we performed this additional robustness check at the city level rather than at the community level.[36] For this additional robustness analysis, we use this treatment intensity definition for Heilongjiang Province based on the identification approach described in our main analysis. Online Appendix Tables B.13–B.15 report the results at the city level. Although we rely on very few observations for this additional robustness analysis, it is evident that both the effect size and the direction of the program effects are consistent with our main estimates based on data from the CHARLS survey. This additional analysis underscores the credibility of the program impacts detected in our main analysis.

---

[34] We also address a potential problem of measurement error at the community-level that may be systematically correlated with unobserved community-level factors in the sample. In the main analysis, we defined our $OfferNRPS_{ct}$ instrument by setting it =1 if at least one person reports having the NRPS in community *c* in time *t*. However, it could be the case that communities with relatively few NRPS participants are correlated in some way, thereby leading to systematic measurement errors in the instrument. To deal with this potential issue, we reran our analyses by omitting communities from our analyses that happen to have the fewest NRPS participants (i.e., communities on the lower end of the distribution of NRPS participants as a fraction of all of the community members). Once we executed the new analysis by omitting communities with very few NRPS participants, the pattern of our findings remained the same. Hence, our results are robust to measurement error coming from communities that misreport NRPS participation. We report the results of this analysis in Online Appendix Tables B.4, B.5, and B.6.

---

[35] We chose the Heilongjiang province because we were able to find city-level announcements on whether specific cities participated in the NRPS program in 2011 and 2013 (the years of the two waves of the CHARLS survey). The region is among the biggest in China, and since identifying variation relies on both time and space, the fact that the region was among the largest was helpful. Other regions had either largely switched to NRPS participation long before 2013, or information about the city available online was not as extensive as it was for Heilongjiang Province.

[36] At the city level, we can compute the treatment intensity for city participation as $city\_participation_{ct}$=(# communities in a city offering $NRPS_{ct}$)/(# total communities in a city)$_{ct}$. Therefore, the real advantage of this robustness check is that we know for sure the number of communities that offer NRPS based on public announcements (the numerator), as opposed to relying on data from the CHARLS to compute that number. The denominator is constant. The downside of this analysis is that the we can only conduct the analysis at the city level (and only for the Heilongjiang Province, for which we obtained data on city/community announcements), which limits the number of observations for this auxiliary analysis.

# 7. Conclusion

This paper provides early estimates of the impact of China's New Rural Pension Scheme (NRPS), first introduced in 2009, by examining the effects of access to pension benefits and retirement on various health domains. Previous studies that explored the effects of retirement on health (Charles 2004; Johnston and Lee 2009; Neuman 2008; Coe and Lindeboom 2008; Coe and Zamarro 2011; Blake and Garrouste 2017; Latif 2013; and Insler 2014) found beneficial physical and mental health impacts, but all of these studies examined the relationships in the context of high-income countries. Our research advances the empirical literature regarding pension programs in developing countries by examining the effects of the NRPS on the health of seniors age 60 and over in rural China, where the economic dynamics resemble the economies of low-income countries.

Our results show that the NRPS had a large and significant short-term positive impact on functional health domains. First, there is striking evidence of the program's positive impact on direct measures of health, including mobility, vision, and self-reported health. Second, there is strong evidence of a beneficial impact on the indirect measures of health, including self-care and usual daily activities. Although there is some evidence of a positive impact on both mental health and social interaction, the effect size estimates for these two domains are imprecisely estimated.

Evidence of the benefits of pension programs that was obtained from developing countries is extremely limited. In China, the CHARLS and CHNS surveys provide a rich dataset that includes data on numerous proxies of health domains. Consequently, we are able to examine how participation in the NRPS retirement program affects numerous previously unexamined domains of health, including mobility, affect, self-care, vision, and social interaction, as well as two disease states. We are among the first studies to provide evidence of the benefits of pension programs in developing countries with respect to functional health domains. These findings support claims that such programs can somewhat reduce the morbidity burden during retirement years.

To gauge potential mechanisms that may drive our results, we also examine how program access influenced specific health behaviors and time use. Our study provides suggestive evidence for some important mechanisms through which access to a pension program likely improves health status. We find that females used the time gained from access to program benefits to engage in longer periods of sleep. In addition, they reduce their incidence of negative health behaviors, such as smoking. For the male sample, although we find a muted positive impact on sleep duration, we do find very large positive program effects in terms of decreased frequency of alcohol consumption. We do not detect any significant impact on access to health care or health care usage for the overall sample.

Our findings have some potentially important implications for the design of retirement policies worldwide. Longevity in developing countries is likely to increase even more in the years to come. Due to the accompanying demographic and socio-economic pressures, pension programs in developing countries are likely to become more prevalent. Our findings point to the potential benefits of introducing partial retirement programs. Our research provides evidence that such public policies can lead to desirable social spillover effects, and they may compress the morbidity burden among the retirement community.

# Figures and Tables

(a) 2009-2010         (b) 2010-2013

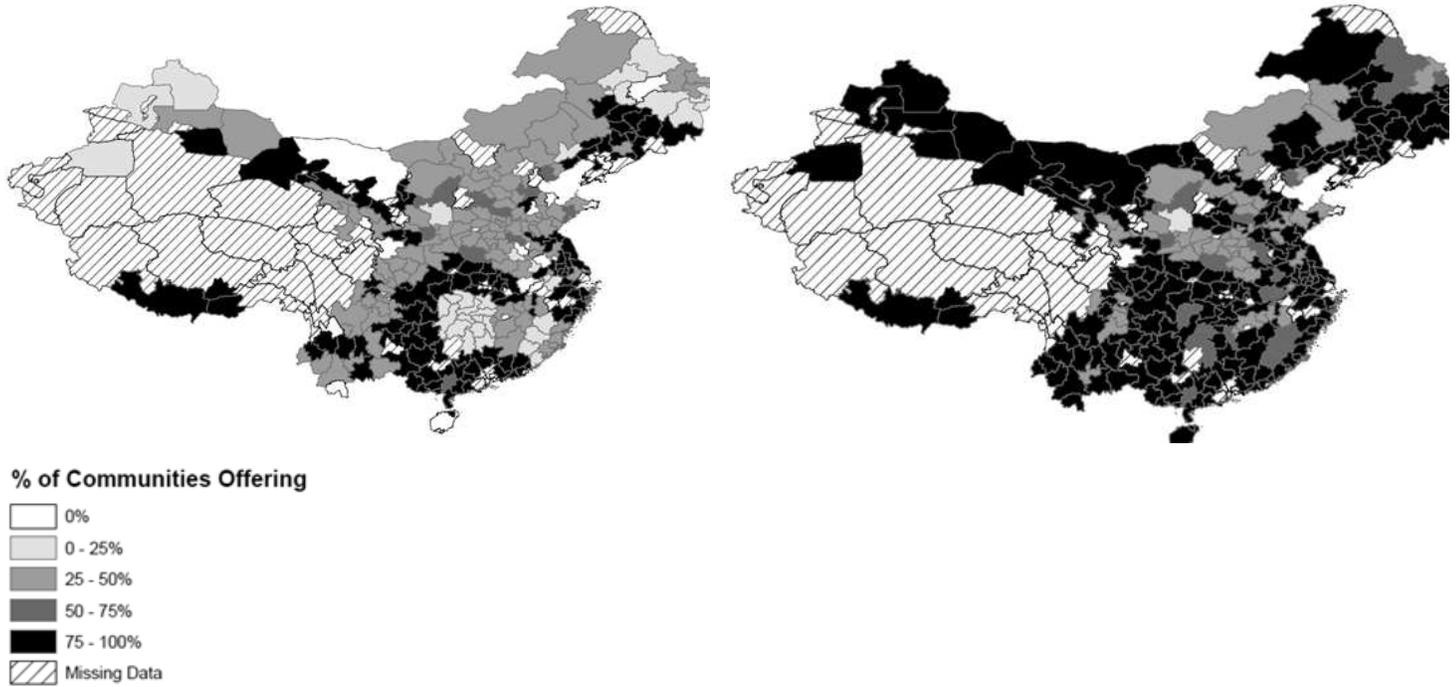

**Fig 1.** Geographic Implementation of NRPS. This figure shows the timely implementation of NRPS. "% of Communities" refers to the percent of communities within the province that offer the NRPS Program.

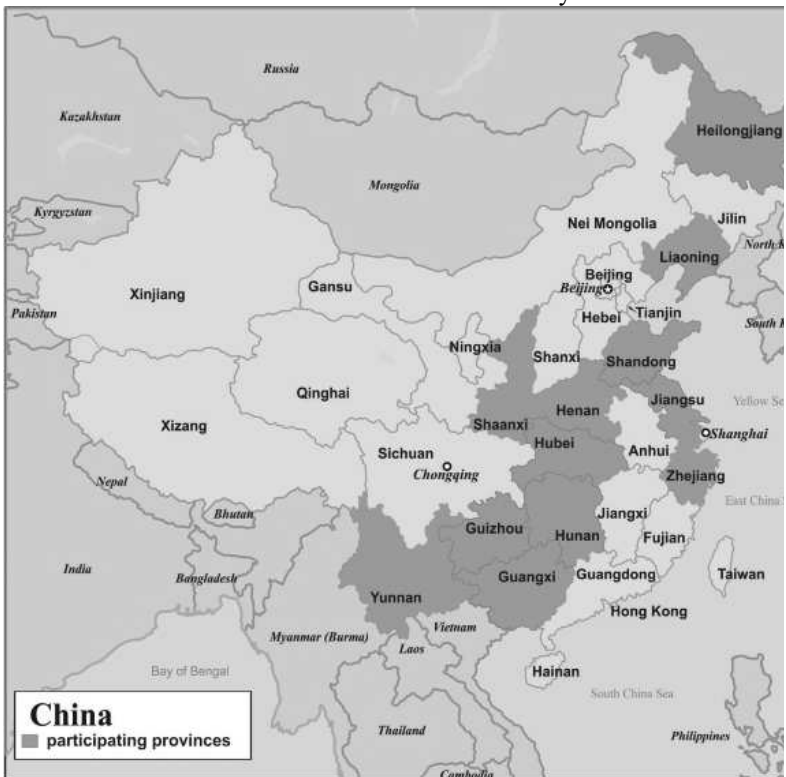

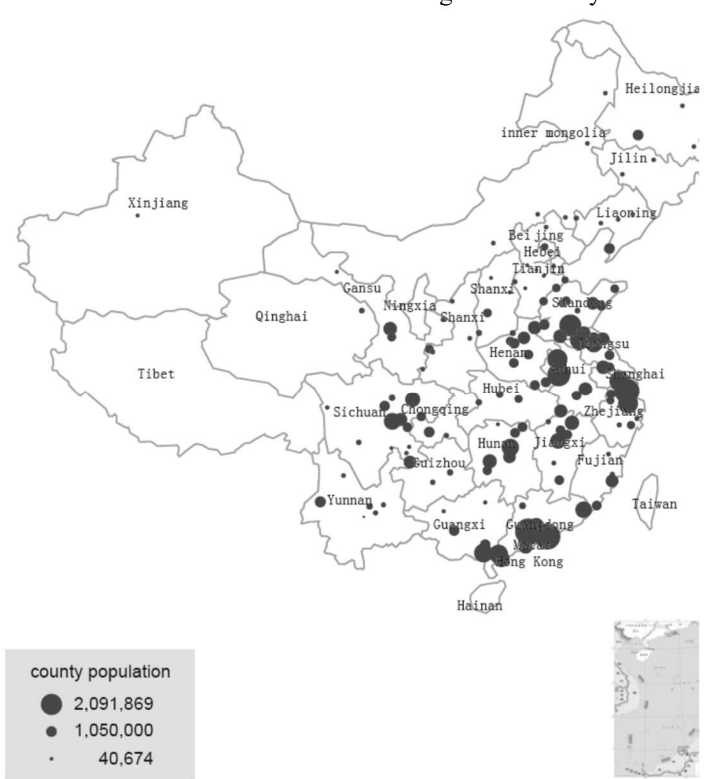

**Fig 2.** Coverage Maps.

**Table 1:** Summary Statistics.

| | Baseline | | | |
|---|---|---|---|---|
| | **Full Sample** | **Participants** | **Non-Participants** | **p-value[a]** |
| *Demographics of Respondents* | | | | |
| Respondent's Age | 59.31 (10.01) | 58.43 (9.68) | 58.44 (10.24) | 0.99 |
| # of Household Residents | 3.74 (1.87) | 3.68 (1.78) | 3.75 (1.88) | 0.04 |
| Percent Female | 0.53 (0.50) | 0.54 (0.50) | 0.53 (0.50) | 0.38 |
| Percent Married | 0.80 (0.40) | 0.81 (0.39) | 0.78 (0.41) | 0.00 |
| Percent With At Least Lower Secondary Education | 0.48 (0.50) | 0.48 (0.50) | 0.46 (0.50) | 0.10 |
| Weekly Work Hours | 45.45 (23.87) | 47.26 (24.07) | 46.89 (22.70) | 0.50 |
| Percent Currently Working | 0.70 (0.46) | 0.70 (0.46) | 0.69 (0.46) | 0.11 |
| Percent Working in Agriculture | 0.72 (0.45) | 0.72 (0.45) | 0.73 (0.45) | 0.49 |
| | | | | |
| *Health Domains and Health Status* | | | | |
| Percent with Difficulty Climbing Flight of Stairs | 0.43 (0.49) | 0.40 (0.49) | 0.43 (0.50) | 0.01 |
| Mobility Index [b] | 0.00 (1.37) | -0.06 (1.35) | 0.00 (1.37) | 0.02 |
| Percent Rarely Feeling Happy | 0.25 (0.43) | 0.17 (0.38) | 0.19 (0.39) | 0.08 |
| Affect Index [b] | 0.00 (1.19) | -0.04 (1.17) | -0.01 (1.24) | 0.27 |
| Short Distance Vision is At Least Good | 0.33 (0.47) | 0.33 (0.47) | 0.33 (0.47) | 0.80 |
| Vision Index [c] | 0.00 (1.19) | -0.03 (1.19) | 0.03 (1.21) | 0.02 |
| Percent with Difficulty Taking Bath | 0.07 (0.26) | 0.06 (0.23) | 0.08 (0.27) | 0.00 |
| Self-Care Index [b] | 0.00 (1.43) | -0.07 (1.34) | 0.06 (1.55) | 0.00 |
| Percent with Difficulty Cleaning House | 0.11 (0.31) | 0.09 (0.29) | 0.11 (0.31) | 0.00 |
| Usual Activities Index [b] | 0.00 (1.47) | -0.06 (1.38) | 0.03 (1.51) | 0.00 |
| Percent with Weekly Contact with Parents or In-Laws | 0.27 (0.44) | 0.30 (0.46) | 0.28 (0.45) | 0.13 |
| Social Interactions Index [c] | 0.00 (1.19) | 0.15 (1.12) | 0.01 (1.16) | 0.00 |
| Percent with At Least Good Self-Reported Health Status | 0.25 (0.43) | 0.27 (0.44) | 0.26 (0.44) | 0.23 |
| | | | | |
| *Health Behavior, Healthcare Utilization and Disease States* | | | | |
| Percent Ever Smoked | 0.41 (0.49) | 0.40 (0.49) | 0.40 (0.49) | 0.98 |
| Percent Smoking Now | 0.25 (0.44) | 0.29 (0.45) | 0.30 (0.46) | 0.39 |
| Percent Consuming Alcohol in Past Year | 0.33 (0.47) | 0.33 (0.47) | 0.33 (0.47) | 0.74 |
| Percent Drinking At Least Once a Week | 0.18 (0.39) | 0.16 (0.37) | 0.16 (0.36) | 0.28 |
| Sleep Duration (hours per night) | 6.28 (1.94) | 6.40 (1.93) | 6.32 (2.00) | 0.07 |
| Number of Doctor Visits | 0.20 (0.40) | 0.20 (0.40) | 0.19 (0.39) | 0.07 |
| Number of Nights Stayed at the Hospital | 0.11 (0.31) | 0.10 (0.29) | 0.09 (0.28) | 0.06 |
| Percent Ever Diagnosed with Diabetes | 0.05 (0.22) | 0.05 (0.22) | 0.04 (0.20) | 0.01 |
| Percent Ever Diagnosed with Hypertension | 0.25 (0.43) | 0.25 (0.43) | 0.22 (0.42) | 0.01 |
| | | | | |
| Observations | 28,034 | 10,011 | 3,680 | |

Notes: Standard deviations are reported in parenthesis. (a) We test the null hypothesis that the difference in participant and non-participant means is equal to 0. (b) Measured in terms of difficulty. Low (or Negative) values denote less difficulty and better mood/emotion. (c) Positively coded where Higher values denote better vision and increased social interaction.

Table 2: ITT and LATE Estimates on Direct Measures of Health.

| | Self-Reported Overall | Mobility | | | | Affect | | | | Vision | | |
|---|---|---|---|---|---|---|---|---|---|---|---|---|
| | At Least Good Health Status (Yes=1) | Diff Jog 1km (Yes=1) | Diff Walk 100m (Yes=1) | Diff Climb Flight of Stairs (Yes=1) | Mobility Index [a] | Everything Mostly an Effort (Yes=1) | Rarely Happy (Yes=1) | Mostly Lonely (Yes=1) | Affect Index [a] | Long Distance Vision | Short Distance Vision | Vision Index [b] |
| | (1) | (2) | (3) | (4) | (5) | (6) | (7) | (8) | (9) | (10) | (11) | (12) |
| **Panel A (ITT):** | | | | | | | | | | | | |
| Offered NRPS * Above60 [c] | 0.022* | -0.024* | -0.013 | -0.057*** | 0.109*** | -0.022* | -0.012 | 0.005 | -0.044 | 0.036** | 0.024 | 0.088** |
| | (0.013) | (0.013) | (0.010) | (0.015) | (0.038) | (0.012) | (0.013) | (0.008) | (0.034) | (0.014) | (0.014) | (0.035) |
| Baseline Mean | 0.252 | 0.543 | 0.172 | 0.427 | 0.000 | 0.167 | 0.248 | 0.079 | 0.000 | 0.353 | 0.328 | 0.000 |
| Controls | Yes | Yes | Yes | Yes | Yes | Yes | Yes | Yes | Yes | Yes | Yes | Yes |
| R-squared | 0.066 | 0.153 | 0.103 | 0.116 | 0.185 | 0.052 | 0.066 | 0.056 | 0.081 | 0.091 | 0.052 | 0.084 |
| Observations | 21,888 | 21,189 | 21,730 | 21,400 | 20,853 | 20,047 | 20,139 | 20,094 | 19,851 | 20,466 | 20,468 | 20,450 |
| **Panel B (TOT):** | | | | | | | | | | | | |
| NRPS Participation * Above60 [d] | 0.50* | -0.52* | -0.028 | -0.127*** | 0.239*** | -0.049* | -0.025 | 0.011 | -0.196 | 0.079** | 0.053 | 0.195** |
| | (0.029) | (0.028) | (0.023) | (0.035) | (0.084) | (0.026) | (0.029) | (0.019) | (0.075) | (0.032) | (0.032) | 0.079 |
| Baseline Mean | 0.252 | 0.543 | 0.172 | 0.427 | 0.000 | 0.167 | 0.248 | 0.079 | 0.000 | 0.353 | 0.328 | 0.000 |
| Controls | Yes | Yes | Yes | Yes | Yes | Yes | Yes | Yes | Yes | Yes | Yes | Yes |
| F-Stat (First Stage) | 246.99 | 251.66 | 249.98 | 251.54 | 251.54 | 249.58 | 251.53 | 251.72 | 251.85 | 247.27 | 247.41 | 274.04 |
| R-squared | 0.065 | 0.152 | 0.102 | 0.110 | 0.182 | 0.051 | 0.065 | 0.056 | 0.080 | 0.088 | 0.051 | 0.082 |
| Observations | 21,888 | 21,189 | 21,730 | 21,400 | 20,853 | 20,047 | 20,139 | 20,094 | 19,851 | 20,466 | 20,468 | 20,450 |

Notes: (a) Measured in terms of difficulty. Low (or Negative) values denote less difficulty and better mobility/mood/emotion. (b) Positively coded where Higher values denote better vision. (c) Our DDD coefficient (Policy instrument interacted with an indicator for being over 60 years old). The control group becomes individuals under the Age of 60 living in eligible communities that didn't offer NRPS between 2011 and 2013. (d) Individual participation instrumented with the policy variable. Also note that Long and Short Distance Vision are indicator variables equal to 1 if either is reported as *At Least Good.* Individual level controls: Marital Status (=1 if Married), Gender (=1 if Female), Education Levels (Base Group is illiterate with no formal education), # of Household Residents. Panel A is estimated using Ordinary Least Squares (OLS) with Community, Year and Community*Year FE. Panel B is estimated using Two-Stage Least Squares (2SLS) with Community, Year and Community*Year FE. Clustered standard errors at the community level reported in parenthesis. *$p < 0.10$, **$p < 0.05$, ***$p < 0.01$.

Table 3: ITT and LATE Estimates on Indirect Measures of Health.

| | Self-Care | | | | Usual Activities | | | | Social Interaction | | | |
|---|---|---|---|---|---|---|---|---|---|---|---|---|
| | Diff Getting Dressed (Yes=1) | Diff Taking Bath (Yes=1) | Diff Eating (Yes=1) | Self-Care Index [a] | Diff Preparing Hot Meals (Yes=1) | Diff Grocery Shopping (Yes=1) | Diff Cleaning House (Yes=1) | Usual Activities Index [a] | Contact with Parents or In-Laws (Yes=1) | Contact with Child over Phone/Email (Yes=1) | Contact with Child in Person/Phone/ Email (Yes=1) | Social Interaction Index [b] |
| | (1) | (2) | (3) | (4) | (5) | (6) | (7) | (8) | (9) | (10) | (11) | (12) |
| **Panel A (ITT): Offered NRPS * Above60[c]** | -0.015** (0.008) | -0.019** (0.009) | -0.020*** (0.006) | -0.145*** (0.050) | -0.017* (0.010) | -0.020* (0.009) | -0.015 (0.010) | -0.105** (0.048) | 0.045*** (0.012) | -0.027 (0.022) | 0.005 (0.009) | 0.031 (0.049) |
| Baseline Mean | 0.056 | 0.071 | 0.029 | 0.000 | 0.096 | 0.098 | 0.106 | 0.000 | 0.270 | 0.523 | 0.909 | 0.000 |
| Controls | Yes | Yes | Yes | Yes | Yes | Yes | Yes | Yes | Yes | Yes | Yes | Yes |
| R-squared | 0.035 | 0.068 | 0.026 | 0.066 | 0.066 | 0.081 | 0.061 | 0.094 | 0.211 | 0.087 | 0.101 | 0.088 |
| Observations | 21,669 | 21,638 | 21,670 | 21,637 | 21,795 | 21,790 | 21,824 | 21,715 | 20,971 | 16,174 | 21,246 | 15,588 |
| **Panel B (TOT): NRPS Participation * Above60[d]** | -0.033** (0.017) | -0.041** (0.020) | -0.044*** (0.013) | -0.320*** (0.112) | -0.038* (0.023) | -0.043** (0.021) | -0.033 (0.022) | -0.231** (0.107) | 0.099*** (0.027) | -0.058 (0.047) | 0.010 (0.020) | 0.067 (0.103) |
| Baseline Mean | 0.056 | 0.071 | 0.029 | 0.000 | 0.096 | 0.098 | 0.106 | 0.000 | 0.270 | 0.523 | 0.909 | 0.000 |
| Controls | Yes | Yes | Yes | Yes | Yes | Yes | Yes | Yes | Yes | Yes | Yes | Yes |
| F-Stat (First Stage) | 251.37 | 250.07 | 251.39 | 250.06 | 250.04 | 248.70 | 248.82 | 250.58 | 254.89 | 240.31 | 245.85 | 243.57 |
| R-squared | 0.034 | 0.066 | 0.019 | 0.062 | 0.065 | 0.079 | 0.060 | 0.091 | 0.206 | 0.086 | 0.101 | 0.087 |
| Observations | 21,669 | 21,638 | 21,670 | 21,637 | 21,795 | 21,790 | 21,824 | 21,715 | 20,971 | 16,174 | 21,246 | 15,588 |

Notes: (a) Measured in terms of difficulty. Low (or Negative) values denote less difficulty and better self-care/usual activities. (b) Positively coded where Higher values denote more social interaction. (c) Our DDD coefficient (Policy instrument interacted with an indicator for being over 60 years old). The control group becomes individuals under the Age of 60 living in eligible communities that didn't offer NRPS between 2011 and 2013. (d) Individual participation instrumented with the policy variable. Individual level controls: Marital Status (=1 if Married), Gender (=1 if Female), Education Levels (Base Group is illiterate with no formal education), # of Household Residents. Panel A is estimated using Ordinary Least Squares (OLS) with Community, Year and Community*Year FE. Panel B is estimated using Two-Stage Least Squares (2SLS) with Community, Year and Community*Year FE. Clustered standard errors at the community level reported in parenthesis. *p< 0.10, **p< 0.05, ***p< 0.01.

| | Health Behavior | | | | | | | Health Utilization | | | | Disease States | |
|---|---|---|---|---|---|---|---|---|---|---|---|---|---|
| | Currently Smoking (Yes=1) | Number of Cigarettes per Day | Drink Alcohol in Past Year (Yes=1) | Drink Regularly in Past Year (Yes=1) | Drink Frequency in Past Year | Under weight (Yes=1) | Sleep Duration (Hours) | Hospital Visit Last Year (Yes=1) | Doctor Visit Last Month (Yes=1) | Hospital Out-of-Pocket (Yuan) | Doctor Out-of-Pocket (Yuan) | Diagnosed w/ Diabetes Past Year (Yes=1) | Diagnosed w/ Hypertension Past Year (Yes=1) |
| | (1) | (2) | (3) | (4) | (5) | (6) | (7) | (8) | (9) | (10) | (11) | (12) | (13) |
| **Panel A (ITT):** | | | | | | | | | | | | | |
| Offered NRPS * Above60 [a] | -0.007 | 0.705** | -0.011 | -0.025*** | -0.136** | -0.007 | 0.105 | 0.014 | -0.018 | 95.824 | 72.338 | 0.001 | 0.001 |
| | (0.010) | (0.331) | (0.011) | (0.009) | (0.056) | (0.008) | (0.076) | (0.010) | (0.012) | (158.902) | (44.753) | (0.003) | (0.005) |
| Baseline Mean | 0.254 | 5.129 | 0.330 | 0.186 | 1.148 | 0.070 | 6.281 | 0.111 | 0.202 | 602.685 | 95.954 | 0.019 | 0.055 |
| Controls | Yes | Yes | Yes | Yes | Yes | Yes | Yes | Yes | Yes | Yes | Yes | Yes | Yes |
| R-squared | 0.344 | 0.336 | 0.263 | 0.217 | 0.238 | 0.051 | 0.013 | 0.038 | 0.036 | 0.063 | 0.058 | | |
| Observations | 18,975 | 17,494 | 21,850 | 20,951 | 20,951 | 16,792 | 20,155 | 21,814 | 21,508 | 21,628 | 21,474 | 18,365 | 18,492 |
| **Panel B (TOT):** | | | | | | | | | | | | | |
| NRPS Participation * Above60 [b] | -0.016 | 1.539** | -0.024 | -0.056*** | -0.304** | -0.015 | 0.231 | 0.030 | -0.039 | 210.812 | 160.146 | 0.002 | 0.003 |
| | (0.023) | (0.709) | (0.024) | (0.021) | (0.128) | (0.018) | (0.170) | (0.021) | (0.026) | (349.876) | (99.155) | (0.007) | (0.011) |
| Baseline Mean | 0.254 | 5.129 | 0.330 | 0.186 | 1.148 | 0.070 | 6.281 | 0.111 | 0.202 | 602.685 | 95.954 | 0.019 | 0.055 |
| Controls | Yes | Yes | Yes | Yes | Yes | Yes | Yes | Yes | Yes | Yes | Yes | Yes | Yes |
| F-Stat (First Stage) | 235.04 | 232.69 | 248.43 | 248.91 | 248.91 | 235.38 | 251.71 | 248.59 | 245.84 | 249.45 | 248.57 | 234.82 | 238.49 |
| R-squared | 0.344 | 0.335 | 0.263 | 0.216 | 0.237 | 0.043 | 0.050 | 0.013 | 0.032 | 0.038 | 0.035 | 0.063 | 0.058 |
| Observations | 18,975 | 17,494 | 21,850 | 20,951 | 20,951 | 16,792 | 20,155 | 21,814 | 21,508 | 21,628 | 21,474 | 18,365 | 18,492 |

Notes: (a) Our DDD coefficient (Policy instrument interacted with an indicator for being over 60 years old). The control group becomes individuals under the Age of 60 living in eligible communities that didn't offer NRPS between 2011 and 2013. (b) Individual participation instrumented with the policy variable. Individual level controls: Marital Status (=1 if Married), Gender (=1 if Female), Education Levels (Base Group is illiterate with no formal education), # of Household Residents. Panel A is estimated using Ordinary Least Squares (OLS) with Community, Year and Community*Year FE. Panel B is estimated using Two-Stage Least Squares (2SLS) with Community, Year and Community*Year FE. Clustered standard errors at the community level reported in parenthesis. *p< 0.10, **p< 0.05, ***p< 0.01.

**Table 5**: LATE Estimates of Retirement on All Outcomes.

| Panel A (Direct Measures of Health): | Self-Reported | Mobility | | | | Affect | | | | Vision | | |
|---|---|---|---|---|---|---|---|---|---|---|---|---|
| | At Least *Good* Health Status (Yes=1) | Diff Jog 1km (Yes=1) | Diff Walk 100m (Yes=1) | Diff Climb Flight of Stairs (Yes=1) | Mobility Index [a] | Everything Mostly an Effort (Yes=1) | Rarely Happy (Yes=1) | Mostly Lonely (Yes=1) | Affect Index [a] | Long Distance Vision | Short Distance Vision | Vision Index [b] |
| | (1) | (2) | (3) | (4) | (5) | (6) | (7) | (8) | (9) | (10) | (11) | (12) |
| Retired (Yes=1) | -3.658 | -0.944 | 0.983 | -0.716 | -0.826 | -7.116** | 25.317*** | -3.097** | 11.947** | -4.074* | 0.507 | -5.395 |
| | (2.707) | (1.773) | (1.246) | (1.879) | (4.365) | (2.814) | (8.851) | (1.315) | (6.054) | (2.315) | (1.869) | (5.155) |
| Baseline Mean | 0.252 | 0.543 | 0.172 | 0.427 | 0.000 | 0.167 | 0.248 | 0.079 | 0.000 | 0.353 | 0.328 | 0.000 |
| Controls | Yes | Yes | Yes | Yes | Yes | Yes | Yes | Yes | Yes | Yes | Yes | Yes |
| F-Stat (First Stage) | 7.49 | 8.22 | 8.37 | 8.32 | 8.14 | 8.24 | 8.28 | 8.09 | 8.13 | 8.31 | 8.30 | 8.31 |
| Beta (First-Stage) | 0.004 | 0.005 | 0.005 | 0.005 | 0.005 | 0.006 | 0.005 | 0.005 | 0.005 | 0.005 | 0.005 | 0.005 |
| SE (First-Stage) | 0.002 | 0.002 | 0.002 | 0.002 | 0.002 | 0.002 | 0.002 | 0.002 | 0.002 | 0.002 | 0.002 | 0.002 |
| Observations | 21,644 | 25,071 | 25,684 | 25,308 | 24,682 | 23,229 | 23,352 | 23,296 | 23,015 | 23,689 | 23,685 | 23,667 |

| Panel B (Indirect Measures of Health): | Self-Care | | | | Usual Activities | | | | Social Interaction | | | |
|---|---|---|---|---|---|---|---|---|---|---|---|---|
| | Diff Getting Dressed (Yes=1) | Diff Taking Bath (Yes=1) | Diff Eating (Yes=1) | Self-Care Index [a] | Diff Preparing Hot Meals (Yes=1) | Diff Grocery Shopping (Yes=1) | Diff Cleaning House (Yes=1) | Usual Activities Index [a] | Contact with Parents or In-Laws (Yes=1) | Contact with Child over Phone/Email (Yes=1) | Contact with Child in Person/Phone/Email (Yes=1) | Social Interaction Index [b] |
| | (1) | (2) | (3) | (4) | (5) | (6) | (7) | (8) | (9) | (10) | (11) | (12) |
| Retired (Yes=1) | -1.748* | -1.590 | -2.525** | -16.580* | -1.598 | -3.276* | 0.697 | -8.209 | -7.278** | -2.318 | -3.549** | -26.506* |
| | (1.059) | (1.189) | (1.191) | (8.585) | (1.192) | (1.721) | (1.060) | (5.981) | (3.396) | (2.991) | (1.762) | (15.834) |
| Baseline Mean | 0.056 | 0.071 | 0.029 | 0.000 | 0.096 | 0.098 | 0.106 | 0.000 | 0.270 | 0.523 | 0.909 | 0.000 |
| Controls | Yes | Yes | Yes | Yes | Yes | Yes | Yes | Yes | Yes | Yes | Yes | Yes |
| F-Stat (First Stage) | 8.36 | 8.36 | 8.36 | 8.36 | 8.47 | 8.53 | 8.48 | 8.47 | 7.96 | 7.25 | 8.41 | 6.67 |
| Beta (First-Stage) | 0.005 | 0.005 | 0.005 | 0.005 | 0.005 | 0.005 | 0.005 | 0.005 | 0.004 | 0.004 | 0.005 | 0.003 |
| SE (First-Stage) | 0.002 | 0.002 | 0.002 | 0.002 | 0.002 | 0.002 | 0.002 | 0.002 | 0.002 | 0.002 | 0.002 | 0.002 |
| Observations | 25,597 | 25,564 | 25,599 | 25,562 | 25,755 | 25,744 | 25,784 | 25,663 | 24,852 | 19,245 | 25,142 | 18,593 |

Notes: (a) Measured in terms of difficulty. Low (or Negative) values denote less difficulty and better mood/emotion. (b) Positively coded where Higher values denote better vision. Also note that Long and Short Distance Vision are indicator variables equal to 1 if either is reported as *At Least Good*. Individual level controls: Age, Age Squared, Marital Status (=1 if Married), Gender (=1 if Female), Education Levels (Base Group is illiterate with no formal education), # of Household Residents. Panel A and Panel B are estimated using Two-Stage Least Squares (2SLS) with Community and Year FE. Clustered standard errors at the community level reported in parenthesis. *p< 0.10, **p< 0.05, ***p< 0.01

Panel C (Health Behavior, Healthcare Utilization and Disease States):

| | **Health Behavior** | | | | | | | | **Health Utilization** | | | **Disease States** | |
|---|---|---|---|---|---|---|---|---|---|---|---|---|---|
| | **Currently Smoking (Yes=1)** | **Number of Cigarettes per Day** | **Drink Alcohol in Past Year (Yes=1)** | **Drink Regularly in Past Year (Yes=1)** | **Drink Frequency in Past Year** | **Underweight (Yes=1)** | **Sleep Duration (Hours)** | **Hospital Visit Last Year (Yes=1)** | **Doctor Visit Last Month (Yes=1)** | **Hospital Out-of-Pocket (Yuan)** | **Doctor Out-of-Pocket (Yuan)** | **Diagnosed w/ Diabetes Past Year (Yes=1)** | **Diagnosed w/ Hypertension Past Year (Yes=1)** |
| | (1) | (2) | (3) | (4) | (5) | (6) | (7) | (8) | (9) | (10) | (11) | (12) | (13) |
| Retired (Yes=1) | -9.826** | -92.007* | 1.062 | 8.648** | 55.552** | -4.387** | -15.564* | 9.533** | 4.176* | 71452.251** | 40738.431** | -0.190 | -0.331 |
| | (4.005) | (52.641) | (1.329) | (3.447) | (22.153) | (1.869) | (8.653) | (3.964) | (2.263) | (29142.579) | (17576.829) | (0.140) | (0.240) |
| Baseline Mean | 0.254 | 5.129 | 0.330 | 0.186 | 1.148 | 0.070 | 6.281 | 0.111 | 0.202 | 602.685 | 95.954 | 0.019 | 0.055 |
| Controls | Yes | Yes | Yes | Yes | Yes | Yes | Yes | Yes | Yes | Yes | Yes | Yes | Yes |
| F-Stat (First Stage) | 6.71 | 6.24 | 8.39 | 8.20 | 8.20 | 7.16 | 8.29 | 8.59 | 8.30 | 8.44 | 8.32 | 7.27 | 7.39 |
| Beta (First-Stage) | 0.005 | 0.005 | 0.005 | 0.005 | 0.005 | 0.004 | 0.005 | 0.005 | 0.005 | 0.005 | 0.004 | 0.004 | 0.004 |
| SE (First-Stage) | 0.002 | 0.002 | 0.002 | 0.002 | 0.002 | 0.002 | 0.002 | 0.002 | 0.002 | 0.002 | 0.002 | 0.002 | 0.002 |
| Observations | 22,794 | 20,876 | 25,812 | 24,920 | 24,920 | 19,873 | 23,344 | 25,833 | 25,459 | 25,610 | 25,442 | 21,801 | 21,963 |

Notes: Individual level controls: Age, Age Squared, Marital Status (=1 if Married), Gender (=1 if Female), Education Levels (Base Group is illiterate with no formal education), # of Household Residents. Panel C is estimated using Two-Stage Least Squares (2SLS) with Community and Year FE. Clustered standard errors at the community level reported in parenthesis. *$p < 0.10$, **$p < 0.05$, ***$p < 0.01$.

**Table 6:** Heterogeneous Treatment Effects Using Male and Female Samples.

| Panel A (Direct Measures of Health): | Self-Reported Overall | Mobility | | | | Affect | | | | Vision | | |
|---|---|---|---|---|---|---|---|---|---|---|---|---|
| | *At Least Good Health Status (Yes=1)* | Diff Jog 1km (Yes=1) | Diff Walk 100m (Yes=1) | Diff Climb Flight of Stairs (Yes=1) | Mobility Index [a] | Everything Mostly an Effort (Yes=1) | Rarely Happy (Yes=1) | Mostly Lonely (Yes=1) | Affect Index [a] | Long Distance Vision | Short Distance Vision | Vision Index [b] |
| | (1) | (2) | (3) | (4) | (5) | (6) | (7) | (8) | (9) | (10) | (11) | (12) |
| **ITT Male Sample:** | | | | | | | | | | | | |
| Offered NRPS * Above60 [c] | 0.016 | -0.037** | -0.007 | -0.041* | -0.085 | -0.022 | 0.002 | -0.001 | -0.037 | 0.045*** | 0.010 | 0.081 |
| | (0.021) | (0.017) | (0.013) | (0.019) | (0.054) | (0.015) | (0.018) | (0.010) | (0.044) | (0.022) | (0.022) | (0.055) |
| Baseline Mean | 0.287 | 0.449 | 0.128 | 0.337 | -0.284 | 0.146 | 0.239 | 0.065 | -0.077 | 0.395 | 0.352 | 0.098 |
| Controls | Yes | Yes | Yes | Yes | Yes | Yes | Yes | Yes | Yes | Yes | Yes | Yes |
| R-squared | 0.054 | 0.119 | 0.066 | 0.090 | 0.142 | 0.033 | 0.062 | 0.059 | 0.073 | 0.083 | 0.047 | 0.088 |
| Observations | 11,049 | 10,789 | 10.984 | 10,838 | 10,646 | 10,022 | 10,037 | 10,023 | 9,934 | 10,189 | 10,195 | 10,185 |
| **ITT Female Sample:** | | | | | | | | | | | | |
| Offered NRPS * Above60 [c] | 0.015 | -0.012 | -0.035** | -0.066*** | -0.143*** | -0.025 | -0.028 | 0.006 | -0.066 | 0.023 | 0.020 | 0.063 |
| | (0.015) | (0.020) | (0.016) | (0.021) | (0.055) | (0.019) | (0.019) | (0.013) | (0.056) | (0.019) | (0.021) | (0.048) |
| Baseline Mean | 0.217 | 0.626 | 0.211 | 0.507 | 0.253 | 0.184 | 0.255 | 0.092 | 0.066 | 0.317 | 0.307 | -0.084 |
| Controls | Yes | Yes | Yes | Yes | Yes | Yes | Yes | Yes | Yes | Yes | Yes | Yes |
| R-squared | 0.046 | 0.121 | 0.110 | 0.095 | 0.160 | 0.046 | 0.048 | 0.030 | 0.063 | 0.073 | 0.031 | 0.058 |
| Observations | 10,737 | 10,302 | 10,644 | 10,644 | 10,108 | 9,942 | 10,220 | 9,989 | 9,834 | 10,190 | 10,186 | 10,178 |
| **TOT Male Sample:** | | | | | | | | | | | | |
| NRPS Participation * Above60 [d] | 0.035 | -0.080** | -0.016 | -0.091* | -0.187 | -0.049 | 0.005 | -0.002 | -0.081 | 0.098** | 0.022 | 0.179 |
| | (0.046) | (0.037) | (0.028) | (0.047) | (0.120) | (0.033) | (0.039) | (0.022) | (0.097) | (0.050) | (0.047) | (0.122) |
| Baseline Mean | 0.287 | 0.449 | 0.128 | 0.337 | -0.284 | 0.146 | 0.239 | 0.065 | -0.077 | 0.395 | 0.352 | 0.098 |
| Controls | Yes | Yes | Yes | Yes | Yes | Yes | Yes | Yes | Yes | Yes | Yes | Yes |
| R-squared | 303.48 | 300.53 | 303.63 | 302.09 | 299.13 | 296.22 | 298.60 | 302.13 | 300.55 | 294.63 | 296.20 | 295.12 |
| F-Stat (First Stage) | 0.053 | 0.119 | 0.066 | 0.087 | 0.141 | 0.033 | 0.062 | 0.059 | 0.073 | 0.075 | 0.045 | 0.082 |
| Observations | 11,049 | 10,789 | 10,984 | 10,838 | 10,646 | 10,022 | 10,037 | 10,023 | 9,934 | 10,189 | 10,195 | 10,185 |
| **TOT Female Sample:** | | | | | | | | | | | | |
| NRPS Participation * Above60 [d] | 0.033 | -0.026 | -0.079** | -0.148*** | -0.318** | -0.056 | -0.062 | 0.014 | -0.147 | 0.052 | 0.046 | 0.143 |
| | (0.035) | (0.044) | (0.037) | (0.050) | (0.125) | (0.043) | (0.042) | (0.030) | (0.126) | (0.043) | (0.047) | (0.110) |
| Baseline Mean | 0.217 | 0.626 | 0.211 | 0.507 | 0.253 | 0.184 | 0.255 | 0.092 | 0.066 | 0.317 | 0.307 | -0.084 |
| Controls | Yes | Yes | Yes | Yes | Yes | Yes | Yes | Yes | Yes | Yes | Yes | Yes |
| F-Stat (First Stage) | 318.63 | 334.12 | 327.40 | 325.01 | 331.53 | 319.57 | 319.96 | 317.30 | 318.99 | 308.80 | 308.98 | 308.16 |
| R-squared | 0.046 | 0.121 | 0.110 | 0.091 | 0.159 | 0.042 | 0.048 | 0.029 | 0.062 | 0.072 | 0.030 | 0.057 |
| Observations | 10,737 | 10,302 | 10,644 | 10,644 | 10,459 | 10,108 | 10,020 | 9,942 | 9,834 | 10,190 | 10,186 | 10,178 |

*Notes:* (a) Measured in terms of difficulty. Low (or Negative) values denote less difficulty and better mood/emotion. (b) Positively coded where Higher values denote better vision. (c) Our DDD coefficient (Policy instrument interacted with an indicator for being over 60 years old). The control group becomes individuals under the Age of 60 living in eligible communities that didn't offer NRPS between 2011 and 2013. (d) Individual participation instrumented with the policy variable. Also note that Long and Short Distance Vision are indicator variables equal to 1 if either is reported as *At Least Good.* Individual level controls: Marital Status (=1 if Married), Education Levels (Base Group is illiterate with no formal education), # of Household Residents. Panel A is estimated using Ordinary Least Squares (OLS) with Community, Year and Community*Year FE. Panel B is estimated using Two-Stage Least Squares (2SLS) with Community, Year and Community*Year FE. Clustered standard errors at the community level reported in parenthesis. *p< 0.10, **p< 0.05, ***p< 0.01.

| Panel B (Indirect Measures of Health): | Self-Care | | | | Usual Activities | | | | Social Interaction | | | |
|---|---|---|---|---|---|---|---|---|---|---|---|---|
| | **Diff Getting Dressed (Yes=1)** | **Diff Taking Bath (Yes=1)** | **Diff Eating (Yes=1)** | **Self-Care Index** [a] | **Diff Preparing Hot Meals (Yes=1)** | **Diff Grocery Shopping (Yes=1)** | **Diff Cleaning House (Yes=1)** | **Usual Activities Index** [a] | **Contact with Parents or In-Laws (Yes=1)** | **Contact with Child over Phone/Email (Yes=1)** | **Contact with Child in Person/Phone/ Email (Yes=1)** | **Social Interaction Index** [b] |
| | (1) | (2) | (3) | (4) | (5) | (6) | (7) | (8) | (9) | (10) | (11) | (12) |
| **ITT Male Sample:** | | | | | | | | | | | | |
| Offered NRPS * Above60 [c] | -0.001 (0.009) | -0.018* (0.010) | -0.012* (0.006) | -0.080 (0.054) | -0.002 (0.012) | -0.014 (0.011) | -0.017 (0.012) | -0.062 (0.056) | 0.034** (0.015) | -0.028 (0.025) | 0.010 (0.011) | 0.051 (0.057) |
| Baseline Mean | 0.050 | 0.063 | 0.026 | -0.042 | 0.094 | 0.070 | 0.092 | -0.082 | 0.281 | 0.524 | 0.904 | -0.011 |
| Controls | Yes | Yes | Yes | Yes | Yes | Yes | Yes | Yes | Yes | Yes | Yes | Yes |
| R-squared | 0.015 | 0.040 | -0.007 | 0.030 | 0.064 | 0.054 | 0.044 | 0.073 | 0.182 | 0.046 | 0.071 | 0.054 |
| Observations | 10,948 | 10,933 | 10,948 | 10,933 | 10,982 | 11,019 | 11,013 | 10,956 | 10,613 | 7,963 | 10,629 | 7,716 |
| **ITT Female Sample:** | | | | | | | | | | | | |
| Offered NRPS * Above60 [c] | -0.032** (0.012) | -0.026* (0.014) | -0.030*** (0.010) | -0.240*** (0.086) | -0.042*** (0.016) | -0.038* (0.015) | -0.021 (0.015) | -0.203*** (0.076) | 0.058*** (0.015) | -0.024 (0.027) | -0.002 (0.012) | 0.021 (0.060) |
| Baseline Mean | 0.061 | 0.078 | 0.031 | 0.037 | 0.098 | 0.123 | 0.119 | 0.072 | 0.261 | 0.522 | 0.913 | 0.009 |
| Controls | Yes | Yes | Yes | Yes | Yes | Yes | Yes | Yes | Yes | Yes | Yes | Yes |
| R-squared | 0.045 | 0.084 | 0.040 | 0.084 | 0.087 | 0.092 | 0.077 | 0.116 | 0.181 | 0.055 | 0.054 | 0.047 |
| Observations | 10,620 | 10,604 | 10,621 | 10,603 | 10,711 | 10,669 | 10,709 | 10,657 | 10,253 | 8,108 | 10,518 | 7,769 |
| **TOT Male Sample:** | | | | | | | | | | | | |
| NRPS Participation * Above60 [d] | -0.002 (0.020) | -0.038* (0.022) | -0.026* (0.014) | -0.175 (0.118) | -0.004 (0.026) | -0.031 (0.023) | -0.038 (0.027) | -0.134 (0.123) | 0.073** (0.034) | -0.061 (0.054) | 0.023 (0.025) | 0.107 (0.121) |
| Baseline Mean | 0.050 | 0.063 | 0.026 | -0.042 | 0.094 | 0.070 | 0.092 | -0.082 | 0.281 | 0.524 | 0.904 | -0.011 |
| Controls | Yes | Yes | Yes | Yes | Yes | Yes | Yes | Yes | Yes | Yes | Yes | Yes |
| F-Stat (First Stage) | 305.35 | 305.21 | 305.91 | 304.64 | 306.32 | 303.55 | 303.11 | 306.21 | 311.38 | 289.10 | 297.33 | 287.59 |
| R-squared | 0.014 | 0.040 | -0.007 | 0.030 | 0.063 | 0.054 | 0.044 | 0.073 | 0.182 | 0.044 | 0.071 | 0.054 |
| Observations | 10,948 | 10,933 | 10,948 | 10,933 | 10,982 | 11,019 | 11,013 | 10,956 | 10,613 | 7,963 | 10,629 | 7,716 |
| **TOT Female Sample:** | | | | | | | | | | | | |
| NRPS Participation * Above60 [d] | -0.072** (0.028) | -0.059* (0.032) | -0.068*** (0.024) | -0.538*** (0.198) | -0.094** (0.037) | -0.085** (0.035) | -0.048 (0.034) | -0.456*** (0.173) | 0.130*** (0.035) | -0.052 (0.059) | -0.005 (0.026) | 0.046 (0.130) |
| Baseline Mean | 0.061 | 0.078 | 0.031 | 0.037 | 0.098 | 0.123 | 0.119 | 0.072 | 0.261 | 0.522 | 0.913 | 0.009 |
| Controls | Yes | Yes | Yes | Yes | Yes | Yes | Yes | Yes | Yes | Yes | Yes | Yes |
| F-Stat (First Stage) | 326.50 | 323.95 | 326.54 | 323.91 | 323.07 | 323.07 | 323.61 | 324.42 | 329.87 | 346.13 | 318.14 | 350.72 |
| R-squared | 0.045 | 0.084 | 0.037 | 0.082 | 0.086 | 0.092 | 0.078 | 0.115 | 0.178 | 0.054 | 0.054 | 0.047 |
| Observations | 10,620 | 10,604 | 10,621 | 10,603 | 10,711 | 10,669 | 10,709 | 10,657 | 10,253 | 8,108 | 10,518 | 7,769 |

Notes: (a) Measured in terms of difficulty. Low (or Negative) values denote less difficulty and better self-care/usual activities. (b) Positively coded where Higher values denote better social interaction. (c) Our DDD coefficient (Policy instrument interacted with an indicator for being over 60 years old). The control group becomes individuals under the Age of 60 living in eligible communities that didn't offer NRPS between 2011 and 2013. (d) Individual participation instrumented with the policy variable. Individual level controls: Marital Status (=1 if Married), Education Levels (Base Group is illiterate with no formal education), # of Household Residents. Panel A is estimated using Ordinary Least Squares (OLS) with Community, Year and Community*Year FE. Panel B is estimated using Two-Stage Least Squares (2SLS) with Community, Year and Community*Year FE. Clustered standard errors at the community level reported in parenthesis. *p< 0.10, **p< 0.05, ***p< 0.0

**Table 6 (Continued)**: Heterogeneous Treatment Effects Using Male and Female Samples.

| Panel C (Health Behavior, Healthcare and Disease States): | Health Behavior | | | | | | | Health Utilization | | | | Disease States | |
|---|---|---|---|---|---|---|---|---|---|---|---|---|---|
| | Currently Smoking (Yes=1) | Number of Cigarettes per Day | Drink Alcohol in Past Year (Yes=1) | Drink Regularly in Past Year (Yes=1) | Drink Frequency in Past Year | Underweight (Yes=1) | Sleep Duration (Hours) | Hospital Visit Last Year (Yes=1) | Doctor Visit Last Month (Yes=1) | Hospital Out-of-Pocket (Yuan) | Doctor Out-of-Pocket (Yuan) | Diagnosed w/ Diabetes Past Year (Yes=1) | Diagnosed w/ Hypertension Past Year (Yes=1) |
| | (1) | (2) | (3) | (4) | (5) | (6) | (7) | (8) | (9) | (10) | (11) | (12) | (13) |
| **ITT Male Sample:** | | | | | | | | | | | | | |
| Offered NRPS * Above60[c] | -0.046* | 0.183 | -0.034* | -0.055*** | -0.333*** | -0.004 | 0.073 | 0.013 | -0.032* | -0.513 | 82.680 | 0.001 | 0.004 |
| | (0.024) | (0.704) | (0.018) | (0.016) | (0.099) | (0.011) | (0.086) | (0.014) | (0.017) | (255.578) | (69.782) | (0.003) | (0.007) |
| Baseline Mean | 0.551 | 12.886 | 0.561 | 0.359 | 2.227 | 0.073 | 6.424 | 0.109 | 0.177 | 666.199 | 82.979 | 0.014 | 0.058 |
| Controls | Yes | Yes | Yes | Yes | Yes | Yes | Yes | Yes | Yes | Yes | Yes | Yes | Yes |
| R-squared | 0.055 | 0.082 | 0.065 | 0.106 | 0.118 | 0.030 | 0.029 | 0.012 | 0.025 | 0.064 | 0.056 | 0.117 | 0.078 |
| Observations | 8,402 | 7,136 | 11,033 | 10,320 | 10,320 | 8,352 | 10,140 | 11,016 | 10,865 | 10,935 | 10,861 | 9,214 | 9,274 |
| **ITT Female Sample:** | | | | | | | | | | | | | |
| Offered NRPS * Above60[c] | -0.025*** | -0.117 | -0.004 | -0.004 | 0.012 | -0.008 | 0.230** | 0.021 | -0.003 | 135.639 | 59.443 | -0.006 | -0.000 |
| | (0.009) | (0.123) | (0.014) | (0.008) | (0.046) | (0.012) | (0.096) | (0.013) | (0.017) | (156.810) | (55.108) | (0.005) | (0.008) |
| Baseline Mean | 0.049 | 0.528 | 0.128 | 0.041 | 0.248 | 0.068 | 6.157 | 0.113 | 0.224 | 547.050 | 107.331 | 0.024 | 0.053 |
| Controls | Yes | Yes | Yes | Yes | Yes | Yes | Yes | Yes | Yes | Yes | Yes | Yes | Yes |
| R-squared | 0.119 | 0.077 | 0.053 | 0.013 | 0.024 | 0.043 | 0.055 | 0.021 | 0.034 | 0.064 | 0.064 | 0.079 | 0.078 |
| Observations | 10,472 | 10,258 | 10,715 | 10,531 | 10,531 | 8,358 | 9,927 | 10,695 | 10,545 | 10,592 | 10,510 | 9,063 | 9,132 |
| **TOT Male Sample:** | | | | | | | | | | | | | |
| NRPS Participation * Above60[d] | -0.101** | 0.398 | -0.074* | -0.123*** | -0.743*** | -0.001 | 0.161 | 0.028 | -0.069* | -1.116 | 180.597 | 0.003 | 0.009 |
| | (0.051) | (1.529) | (0.039) | (0.037) | (0.227) | (0.024) | (0.189) | (0.029) | (0.036) | (556.394) | (152.653) | (0.006) | (0.014) |
| Baseline Mean | 0.551 | 12.886 | 0.561 | 0.359 | 2.227 | 0.073 | 6.424 | 0.109 | 0.177 | 666.199 | 82.979 | 0.014 | 0.058 |
| Controls | Yes | Yes | Yes | Yes | Yes | Yes | Yes | Yes | Yes | Yes | Yes | Yes | Yes |
| F-Stat (First Stage) | 210.33 | 200.58 | 305.17 | 306.80 | 306.80 | 281.06 | 297.11 | 304.92 | 301.07 | 301.94 | 301.11 | 293.09 | 295.11 |
| R-squared | 0.056 | 0.083 | 0.061 | 0.102 | 0.114 | 0.030 | 0.025 | 0.005 | 0.022 | 0.064 | 0.054 | 0.116 | 0.077 |
| Observations | 8,402 | 7,136 | 11,033 | 10,320 | 10,320 | 8,352 | 10,140 | 11,016 | 10,865 | 10,935 | 10,861 | 9,214 | 9,274 |
| **TOT Female Sample:** | | | | | | | | | | | | | |
| NRPS Participation * Above60[d] | -0.055*** | -0.262 | -0.010 | -0.010 | 0.026 | -0.019 | 0.518** | 0.048 | -0.008 | 304.790 | 135.011 | -0.014 | -0.001 |
| | (0.020) | (0.278) | (0.031) | (0.019) | (0.104) | (0.027) | (0.218) | (0.030) | (0.039) | (352.850) | (124.929) | (0.012) | (0.017) |
| Baseline Mean | 0.049 | 0.528 | 0.128 | 0.041 | 0.248 | 0.068 | 6.157 | 0.113 | 0.224 | 547.050 | 107.331 | 0.024 | 0.053 |
| Controls | Yes | Yes | Yes | Yes | Yes | Yes | Yes | Yes | Yes | Yes | Yes | Yes | Yes |
| F-Stat (First Stage) | 320.64 | 323.69 | 321.83 | 321.29 | 321.29 | 276.56 | 311.74 | 323.40 | 317.55 | 322.98 | 321.19 | 306.43 | 313.57 |
| R-squared | 0.115 | 0.077 | 0.054 | 0.012 | 0.024 | 0.040 | 0.051 | 0.020 | 0.035 | 0.063 | 0.064 | 0.079 | 0.078 |
| Observations | 10,472 | 10,258 | 10,715 | 10,531 | 10,531 | 8,358 | 9,927 | 10,695 | 10,545 | 10,592 | 10,510 | 9,063 | 9,132 |

Notes: (a) Our DDD coefficient (Policy instrument interacted with an indicator for being over 60 years old). The control group becomes individuals under the Age of 60 living in eligible communities that don't offer NRPS between 2011 and 2013. (b) Individual participation instrumented with the policy variable. Individual level controls: Marital Status (=1 if Married), Gender (=1 if Female), Education Levels (Base Group is illiterate with no formal education), # of Household Residents. Panel A is estimated using Ordinary Least Squares (OLS) with Community, Year and Community*Year FE. Panel B is estimated using Two-Stage Least Squares (2SLS) with Community, Year and Community*Year FE. Clustered standard errors at the community level reported in parenthesis. *p< 0.10, **p< 0.05, ***p< 0.01